\documentclass[journal]{IEEEtran}

\usepackage{amsmath,amsfonts,amssymb}
\usepackage{algorithmic}
\usepackage{algorithm}
\usepackage{array}
\usepackage[caption=false,font=normalsize]{subfig}
\usepackage{textcomp}
\usepackage{stfloats}
\usepackage{url}
\usepackage{verbatim}
\usepackage{graphicx}
\usepackage{cite}
\usepackage{amsthm}
\usepackage{xcolor}
\usepackage{utfsym}
\usepackage{dcolumn}
\usepackage{orcidlink}
\usepackage{hyperref}
\hypersetup{
    colorlinks=true,
    linkcolor=blue,
    filecolor=magenta,      
    urlcolor=cyan,
    citecolor=cyan,
    }

\newtheorem{thm}{Theorem} 

\newtheorem{assum}{Assumption}
\newtheorem{remark}{Remark}

\AddToHook{env/thebibliography/begin}{\color{black}\hypersetup{allcolors=black}}

\AtBeginDocument{%
  \setlength{\abovedisplayskip}{3pt plus 1pt minus 1pt}%
  \setlength{\belowdisplayskip}{3pt plus 1pt minus 1pt}%
  \setlength{\abovedisplayshortskip}{2pt plus 1pt minus 1pt}%
  \setlength{\belowdisplayshortskip}{2pt plus 1pt minus 1pt}%
}
\allowdisplaybreaks
\begin{document}

\title{Boundary Stabilization of Mixed Traffic Flow With Connected Automated Vehicles}

\author{Yihuai~Zhang, Ziming Wang, Lu Liu*
\thanks{This work was supported by the Research Grants Council of the Hong
Kong Special Administrative Region of China under Project CityU/11212225. (Corresponding author: Lu Liu.)}
\thanks{Yihuai Zhang and Lu Liu are with the Department of Mechanical Engineering, City University of Hong Kong, Hong Kong, China.~{E-mails: \tt yihuai.zhang@cityu.edu.hk; luliu45@cityu.edu.hk}}
\thanks{Ziming Wang is with the Thrust of Artificial Intelligence, The Hong Kong University of Science and Technology (Guangzhou), Guangzhou, China. E-mail:{\tt~zwang216@connect.hkust-gz.edu.cn}}}%

\maketitle

\begin{abstract}
This paper addresses boundary stabilization of stop-and-go oscillations in mixed traffic containing human-driven vehicles (HVs) and connected automated vehicles (CAVs). A penetration-dependent fundamental diagram is constructed from vehicle-class headways and effective spacing and embedded in an Aw--Rascle--Zhang (ARZ) model. Backstepping yields full-state and output-feedback controllers for the nominal linearized system. Exponential stability and finite-time convergence are established for the linearized system. Closed-loop simulations over CAV penetration rates from 0 to 1 show that increasing penetration raises the equilibrium speed and flow and reduces the convergence time for both the state-feedback and output-feedback controllers. A proportional-integral (PI) controller benchmark, input magnitude and rate constraints, physical sensor noise, and a sensitivity analysis are also considered to assess the performance of the controllers. We further illustrate the effectiveness of the proposed control methods and the simulation framework through a fundamental diagram calibration using the TGSIM dataset. These results quantify the traffic performance changes and identify the limits of the proposed method.
\end{abstract}


\section{Introduction}\label{sec:introduction}
\IEEEPARstart{F}{reeway} traffic congestion reduces throughput, increases fuel consumption and emissions, and raises the risk of accidents~\cite{siri2021freeway}. There are many causes of traffic congestion, including bottlenecks, lane changes, and  {differences in} human driver  {behavior}~\cite{treiber2013traffic}. The resulting oscillations can propagate upstream  {over} long distances and persist for long  {periods}. A variety of traffic control methods have been proposed to mitigate these oscillations, such as feedback control~\cite{papageorgiou1991alinea,karafyllis2019feedback,zhang2017stochastic}, optimal control~\cite{gomes2006optimal,goatin2016speed,delle2017traffic}, learning-based control~\cite{belletti2017expert}, and vehicular formation control~\cite{wang2025adaptive,wang2026event,wang2026safe}. At the implementation level, these methods can be  {realized} through variable speed limits (VSLs)~\cite{yu2018varying} and ramp metering~\cite{papageorgiou2002freeway}. With the development of connected automated vehicles (CAVs), CAVs can also be used to stabilize traffic flow or improve throughput~\cite{TalebpourInfluenceConnected2016}. CAVs are expected to change freeway traffic operations through shorter response times, vehicle-to-vehicle communication, and cooperative longitudinal control. Their benefits, however, will emerge during a long transition period in which CAVs and human-driven vehicles (HVs) share the same road. In this mixed regime, traffic stability and capacity depend not only on the total density, but also on the CAV penetration rate and the different spacing policies of HVs and CAVs. Prior studies have shown that controlled  {automated vehicles} can damp stop-and-go waves even at low penetration levels~\cite{SternDissipationStopandgo2018}. These observations motivate control-oriented traffic models that explicitly connect mixed-vehicle composition with traffic stability.

Macroscopic partial differential equation (PDE) models provide a natural framework for describing the spatiotemporal evolution of freeway traffic. Classical first-order models, represented by the
Lighthill-Whitham-Richards (LWR) model~\cite{lighthill1955kinematic,richards1956shock}, describe density
evolution through a conservation law and a fundamental diagram. Second-order models, such as the Payne-Whitham model~\cite{payne1971model,whitham2011linear}, introduce velocity  {dynamics but} may exhibit
nonphysical behavior. The Aw-Rascle-Zhang (ARZ) model overcomes this limitation by describing density
and velocity through a hyperbolic PDE system with anisotropic traffic propagation~\cite{aw2000resurrection,zhang2002non}. Because stop-and-go waves in congested traffic involve both density and velocity perturbations, the ARZ model is particularly suitable for boundary stabilization problems in freeway traffic~\cite{yu2019traffic}.

A key modeling  {component} in mixed traffic is the fundamental diagram. Classical diagrams are usually calibrated for pure HV traffic and therefore do not directly represent the shorter headways, smaller gaps, and cooperative following behavior enabled by CAVs. Recent studies have extended
fundamental diagram modeling to automated or mixed traffic. Ghiasi et al.~\cite{ghiasi2017mixed} developed a mixed-traffic capacity model based on stochastic headway distributions and CAV platooning intensity. Shi and Li~\cite{shi2021constructing} constructed fundamental diagrams for automated-vehicle traffic and showed that headway settings affect both capacity and stability. Yao et al.~\cite{yao2019stability} derived a mixed CAV-HV fundamental diagram and {a} stability condition under
different penetration rates. Zhou and Zhu~\cite{zhou2020modeling} further modeled the mixed HV-CAV fundamental diagram by accounting for CAV penetration and platooning intensity. These studies clarify how CAV penetration changes  {the} capacity and equilibrium states of  {the} traffic system, but they mainly use the fundamental diagram for traffic-flow characterization rather than for controller design in the mixed-traffic
ARZ model. Table~\ref{summary} summarizes related studies and their contributions to mixed-traffic fundamental diagram modeling.
\begin{table}[tbp]
\centering
\caption{Fundamental-diagram modeling for mixed traffic}\small\setlength{\tabcolsep}{3pt}
\begin{tabular}{ccccc}
    \hline
    Article & Vehicle types & Stability & Platoon intensity &  PDE-type \\
    \hline
    \cite{shi2021constructing} & AVs/HVs & \usym{2713} & \usym{2717} & \usym{2717}\\
    \cite{ye2018modeling} & CAVs/HVs & \usym{2717} & \usym{2717} & \usym{2717}\\
    \cite{yao2019stability} & CAVs/HVs & \usym{2713} & \usym{2717} & \usym{2717}\\
    \cite{chang2020analysis} & ICV platoon/HVs & \usym{2713} & \usym{2717} & \usym{2717}\\
    \cite{zhou2020modeling} & CAVs/HVs & \usym{2713} & \usym{2713} & \usym{2717}\\
    \cite{ghiasi2017mixed} & CAVs/HVs & \usym{2717} & \usym{2713} &\usym{2717}\\ 
    Ours & CAVs/HVs & \usym{2713} & \usym{2717} &\usym{2713}\\
    \hline
\end{tabular} 
\label{summary}
\end{table}
In addition to mixed-traffic fundamental diagram modeling, Burkhardt et
al.~\cite{BurkhardtStopandgoSuppression2021} developed a backstepping-based boundary controller for suppressing stop-and-go waves in a two-class ARZ model. They  {used} an extended ARZ model, represented as a $4\times4$ hyperbolic PDE system, to describe two-class traffic dynamics. The same modeling framework was later extended to mixed-autonomy traffic systems~\cite{ZhangMeansquareExponential2024,ZhangEventTriggeredRegulation2026,zhang2024event}. However, the fundamental diagrams in these studies are not explicitly constructed from vehicle-dependent headways and CAV penetration rates.

Boundary control is  {well suited} for freeway regulation because traffic management actions are often applied at road boundaries, such as VSLs and ramp metering. PDE backstepping has become a powerful method for stabilizing heterodirectional hyperbolic systems~\cite{krstic2008boundary,VazquezBacksteppingBoundary2011,anfinsen2019adaptive}. For traffic applications, Yu and Krstic~\cite{yu2019traffic} designed boundary feedback controllers for the ARZ model and proved exponential stability and finite-time convergence in  {the} congested regime. Related studies have extended
ARZ-based freeway control to VSL control~\cite{yu2018varying}, two-lane traffic~\cite{YuOutputFeedback2021}, and mixed ACC/manual traffic through time-gap manipulation~\cite{bekiaris2020pde}. Delay-compensated PDE control has also been developed for connected/automated traffic~\cite{qi2022delay}. These works provide important foundations for traffic control, but they do not directly address the case in which  {a penetration-dependent} fundamental diagram is embedded into the ARZ model.

Motivated by these gaps, this paper develops a boundary control framework for mixed traffic composed of HVs and CAVs. We first construct a mixed fundamental diagram from  {the} time headways and effective spacing {s} of different vehicle types, then embed this diagram into the ARZ PDE model, and finally design full-state and output-feedback boundary controllers for the resulting congested  {mixed-traffic} system. The paper focuses on  {a general one-lane freeway segment} and studies stabilization around a congested equilibrium. The main contributions are summarized as follows.
  \begin{itemize}
      \item A mixed HV-CAV fundamental diagram is  {developed} by incorporating the CAV penetration rate,  {heterogeneous} time headways, and effective spacing into the density-flow and velocity-density  {relations}. The diagram is embedded into the ARZ model to obtain  {the resulting PDE model} of mixed traffic.
      \item Backstepping is applied to the nominal linearized mixed-traffic model through full-state and observer-based boundary feedback. The ideal downstream regulation channel is implemented by VSL actuation.
      \item Extensive simulations quantify the effects of the proposed control methods and CAV penetration rates, compared with a benchmark PI controller. A calibrated fundamental diagram calibrated using real traffic data is used to illustrate the effectiveness of the proposed control methods and the simulation framework. 
    \end{itemize}

\section{Problem Statement}
\label{sec:problem}
In this section, we formulate the problem and develope the mixed fundamental diagram and the coordinate transformation for the backstepping design. The major notations to be used are summarized in Table~\ref{tab:notations}.
\begin{table}
    \centering
    \caption{Major notations}\label{tab:notations}
    \begin{tabular}{ll}
    \hline
        Notation & Description\\
        \hline
        $x$ & Spatial coordinate along the freeway segment\\
        $t$ & Time variable\\
        $\rho(x,t),\rho^*$ & Traffic density and its equilibrium\\
        $v(x,t),v^*$ & Traffic speed and its equilibrium\\
        $\tilde{\rho}$ & Deviation of traffic density from its equilibrium\\
        $\tilde{v}$ & Deviation of traffic speed from its equilibrium\\
        $V(\rho)$ & Fundamental diagram \\
        $v-V(\rho)$ & Speed departure from equilibrium\\
        $\tau$ & Adaptation time \\
        $P_{\text{CAV}}$ & Penetration rate of CAVs\\
        $h_0, h_1$ & Headways for HVs and CAVs\\
        $l$ & Average vehicle length\\
        $s_g^0, s_g^1$ & Minimum standstill gaps for HVs and CAVs\\
        $\rho_{\text{mix}}$ & Mixed traffic density\\
        $h_{\text{mix}}$ & Mixed time headway \\
        $l_{\text{eff}}$ & Effective vehicle length\\
        $\tilde{w}$ & Traffic states in the Riemann coordinates\\
        $\bar{w}$ & Scaled coordinate of traffic states for backstepping\\
        $U_b(t)$ & Physical boundary control input\\
        $U(t)$ & normalized boundary control input\\
        \hline
    \end{tabular}
\end{table}
\subsection{ARZ Model and Mixed Fundamental Diagram}\label{sec:arz-fd}
We use the ARZ model with relaxation to describe the traffic density and velocity on a single freeway segment~\cite{aw2000resurrection,zhang2002non}. The dynamics are given by
\begin{equation}
    \begin{aligned}
        \partial_t \rho + \partial_x (\rho v) &= 0,\\
        \partial_t (v - V(\rho)) + v \partial_x (v - V(\rho)) &= \frac{V(\rho) - v}{\tau},
    \end{aligned}
\end{equation}
where $\rho(x,t)$ is the traffic density and $v(x,t)$ is the traffic velocity. The spatiotemporal domain is given by $x \in [0, L]$ and $t \in \mathbb{R}^+$,  where $L$ denotes the length of the freeway segment. Here, $\tau>0$ denotes the adaptation time, characterizing the relaxation of traffic speed toward the equilibrium speed associated with the current density. The function $V(\rho)$ is the fundamental diagram relating traffic density to velocity.  The deviation $w_{\rm nl}=v-V(\rho)$ measures speed departure from equilibrium at the current density. Along a vehicle trajectory, $\partial_t w_{\rm nl}+v\partial_x w_{\rm nl}=-w_{\rm nl}/\tau$ relaxes this departure toward zero.
 However,  fundamental diagrams calibrated for pure-HV traffic cannot directly represent mixed traffic flow. Although CAVs and HVs may occupy the same density range, they  generally follow different headway policies.  To construct a fundamental diagram for a road segment containing both vehicle types, we adopt the following assumptions~\cite{zhou2020modeling}.
\begin{assum}\label{assum1}
    Four following behaviors,  {including CAV-following-CAV, CAV-following-HV, HV-following-CAV, and HV-following-HV}, may occur in mixed traffic~\cite{zhou2020modeling,yao2023analysis}. For tractability, these behaviors are represented by CAV-following  {headway} and HV-following  {headway}. We use $\{h_{0}, h_{1}\}$ to denote the two headways, where $h_0$ is the  {HV} headway and $h_1$ is the  {CAV} headway. The headway is measured from the rear bumper of the leading vehicle to the front bumper of the following vehicle. A  {constant time-headway} (CTH) policy is  {assumed} for  {both} HVs and CAVs.
\end{assum}
\begin{remark}
    This aggregation treats vehicle ordering through an  {average-spacing} relation. It therefore captures  {the effect of the penetration rate} on the macroscopic fundamental diagram, but it does not model stochastic leader-follower sequences, lane changing, or platoon-size constraints. These effects are  {beyond the scope of} the boundary stabilization framework developed in this paper.
\end{remark}

\begin{assum}\label{assum3}
The CAV penetration rate is defined as
 {\begin{align}
    P_{\text{CAV}} =  \frac{\sum_{i=1}^{N}A_i}{N},
\end{align}}
where $N$ is the total number of vehicles on the road section and $A_i$ denotes the vehicle type of the $i$th vehicle:
\begin{align}
    A_i=\begin{cases}
        1,&\text{if the }i\text{th vehicle is a CAV},\\
        0,&\text{if it is an HV}.
    \end{cases}
\end{align}
\end{assum}
For the nominal model, $P_{\rm CAV}$ is known, spatially uniform and constant in time. The follower-class closure neglects ordering and lane-change dynamics.
The CAV penetration rate enters the diagram through spacing. Starting from the standard density-spacing relation~\cite{treiber2013traffic},
\begin{align}\label{eq:density-spcing}
    \rho = \frac{1}{s},
\end{align}
where $\rho$ is {the} traffic density and $s$ is {the} distance headway.
Let $k\in\{0,1\}$ denote the vehicle type, with $k=0$ for HVs and $k=1$ for CAVs. Under the CTH policy, the spacing for either vehicle type is
\begin{align}
    s^{(k)} = l + s_g^k + h_k v,
    \label{eq:class-spacing}
\end{align}
where $l$ is the average vehicle length, and $s_g^k$ and $h_k$ are the minimum standstill gap and time headway of type-$k$ vehicles, respectively. For HVs, this relation is the CTH approximation to the equilibrium spacing of the intelligent driver model~(IDM)~\cite{kesting2010enhanced}. The corresponding traffic flow-density relation is
\begin{align}
    Q_k(\rho_k) = \begin{cases}
        \rho_k v_f, & 0\leq\rho_k\leq\rho_k^c,\\[2pt]
        \dfrac{1-\rho_k(l+s_g^k)}{h_k}, & \rho_k^c<\rho_k\leq\rho_k^{\max},
    \end{cases}
    \label{eq:class-flow-density}
\end{align}
where $\rho_k$ is the density of a homogeneous stream of type-$k$ vehicles, $\rho_k^c=1/(h_kv_f+l+s_g^k)$ is its critical density, and $\rho_k^{\max}=1/(l+s_g^k)$ is its jam density. The velocity-density relation is
\begin{align}
    V_k(\rho_k) = \begin{cases}
        v_f, & 0\leq\rho_k\leq\rho_k^c,\\[2pt]
        \dfrac{1-\rho_k(l+s_g^k)}{\rho_k h_k}, & \rho_k^c<\rho_k\leq\rho_k^{\max}.
    \end{cases}
    \label{eq:class-speed-density}
\end{align}

The mixed diagram is obtained by averaging the two spacing policies in a locally homogeneous congested stream, where HVs and CAVs share a common equilibrium speed $v$. Neglecting end effects, we associate each vehicle with the gross spacing $s^{(A_i)}$ defined in \eqref{eq:class-spacing}. The mixed density is the reciprocal of the vehicle-averaged spacing:
\begin{align}
    \rho_{\text{mix}} = \frac{N}{\sum_{i=1}^{N} s^{(A_i)}},
    \label{eq:mixed-density-average}
\end{align}
where $A_i=0$ identifies an HV and $A_i=1$ identifies a CAV. Substituting the two spacing policies and using $P_{\text{CAV}}=N^{-1}\sum_{i=1}^{N}A_i$ gives
\begin{align}
    \begin{aligned}
        \rho_{\text{mix}}
        &= \frac{N}{Nl+\sum_{i=1}^{N}(s_g^{A_i}+h_{A_i}v)}\\
        &= \frac{1}{l_{\text{eff}}+h_{\text{mix}}v},
    \end{aligned}
    \label{density}
\end{align}
where the mixed time headway is $h_{\text{mix}} = (1-P_{\text{CAV}})h_0+P_{\text{CAV}}h_1$ and effective length is $l_{\text{eff}} = l+(1-P_{\text{CAV}})s_g^0+P_{\text{CAV}}s_g^1$. Thus, both vehicle types contribute to the mean standstill gap. Hereafter, $\rho(x,t)$ denotes the total mixed density $\rho_{\text{mix}}(x,t)$. Combining the congested relation in \eqref{density} with the common free-flow speed $v_f$ yields
\begin{align}
    V_{\text{mix}}(\rho) = \begin{cases}
        v_f, & 0\leq\rho\leq\rho^c_{\text{mix}},\\[2pt]
        \dfrac{1/\rho-l_{\text{eff}}}{h_{\text{mix}}},
        & \rho^c_{\text{mix}}<\rho\leq\rho^{\max}_{\text{mix}},
    \end{cases}
    \label{mix-fun}
\end{align}
and the flow-density relation $Q_{\text{mix}}(\rho)=\rho V_{\text{mix}}(\rho)$ is
\begin{align}
    Q_{\text{mix}}(\rho) = \begin{cases}
        \rho v_f, & 0\leq\rho\leq\rho^c_{\text{mix}},\\[2pt]
        \dfrac{1-\rho l_{\text{eff}}}{h_{\text{mix}}},
        & \rho^c_{\text{mix}}<\rho\leq\rho^{\max}_{\text{mix}}.
    \end{cases}
    \label{eq:mixed-flow-density}
\end{align}
Here, $\rho^c_{\text{mix}}=1/(v_fh_{\text{mix}}+l_{\text{eff}})$ is the critical density, and $\rho^{\max}_{\text{mix}}=1/l_{\text{eff}}$ is the jam density. At $P_{\text{CAV}}=0$ and $P_{\text{CAV}}=1$, the mixed diagram reduces to the corresponding HV and CAV diagrams in \eqref{eq:class-flow-density} and \eqref{eq:class-speed-density}.

\begin{figure}[tbp]
    \centering
    \subfloat[Flow-density]{\includegraphics[width = 0.48\linewidth]{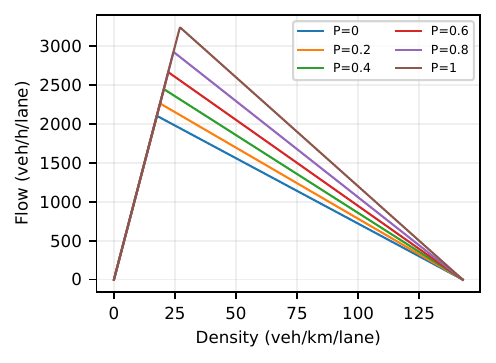}}
    \subfloat[Speed-density]{\includegraphics[width = 0.48\linewidth]{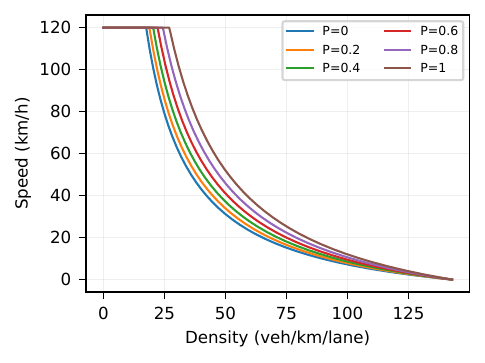}}
    \caption{Fundamental diagram of mixed-traffic scenarios}
    \label{Fundamental diagram of mixed-traffic}
\end{figure}

Table~\ref{Parameters of HVs and CAVs} lists the nominal parameters. For HVs, we use a standstill gap $s_g^0=2$ m~\cite{treiber2000congested} and a time headway $h_0=1.5$ s~\cite{kesting2010enhanced}. For CAVs, $s_g^1=2$ m is adopted as a positive standstill-gap setting, and $h_1=0.9$ s follows a CACC time-gap setting tested in~\cite{milanes2013cooperative}. Both vehicle types have length $l=5$ m and free-flow speed $v_f=120$ km/h. Consequently, $l_{\text{eff}}=7$ m is independent of penetration, while $h_{\text{mix}}=(1.5-0.6P_{\text{CAV}})$ s decreases with penetration. The capacity $Q_{\max}=v_f/(v_fh_{\text{mix}}+l_{\text{eff}})$ therefore increases because of the reduced mean time headway. Figure~\ref{Fundamental diagram of mixed-traffic} illustrates the flow-density and velocity-density relations of the mixed-traffic model.

\begin{table}[tbp]
    \caption{Parameters of HVs and CAVs}\label{Parameters of HVs and CAVs}
    \begin{center}
    \begin{tabular}{ c c c }
    \hline
    \textbf{Parameters} & \textbf{HVs} & \textbf{CAVs}\\
    \hline
    Minimum gap $s_g$ (m) & 2 & 2 \\
    Headway $h$ (s) & 1.5 & 0.9 \\
    Vehicle length $l$ (m) & 5 & 5 \\
    Maximum speed $v_f$ (km/h) & 120 & 120\\
    \hline
    \end{tabular}
    \end{center}
\end{table}

We now place this diagram inside the ARZ dynamics. The resulting PDE system is written in quasi-linear form as
\begin{equation}
    \begin{aligned}
    \partial_t \rho + v \partial_x\rho + \rho \partial_x v &= 0,\\
    \partial_t v + \left( v + \rho V_{\text{mix}}'(\rho) \right) \partial_x v  &=\frac{V_{\text{mix}}(\rho) - v }{\tau},
    \end{aligned}
\end{equation}
where $V_{\text{mix}}'(\rho)=-1/(\rho^2h_{\text{mix}})$. Let $\rho^*$ and $v^*$ denote the equilibrium density and velocity. At equilibrium, $\partial_x v^*=\partial_t v^*=\partial_x\rho^*=\partial_t\rho^*=0$, so the dynamics reduce to
\begin{align}
    V_{\text{mix}}(\rho^*) = v^*.
\end{align}
The inlet flow is held at $q^* = \rho^* v^*$, and the designed control input acts at the downstream boundary as an outlet speed-regulation term. This representation is consistent with a VSL-type command near the segment exit.
The boundary conditions are
\begin{equation}
    \begin{aligned}
    \rho(0,t) = \frac{q^*}{v(0,t)},
    v(L,t) = \frac{q^*}{\rho(L,t)} + U_b(t),
    \end{aligned}
\end{equation}
where $U_b(t)$ is the physical boundary input with speed units. It is distinct from the normalized design input $U(t)$ introduced below.

\subsection{Transformation of Mixed-traffic ARZ Model}\label{sec:transformation}\label{sec:linear-model}
We first linearize around the congested equilibrium $(\rho^*,v^*)$, with $v^*=V_{\rm mix}(\rho^*)$, and define the density and speed perturbations $\tilde\rho=\rho-\rho^*$ and $\tilde v=v-v^*$.
Set
\begin{equation}
    \begin{aligned}
    S&=-\rho^*V'_{\rm mix}(\rho^*),& \mu&=S-v^*>0,\\
    r_{\mathrm b}&=-\frac{\mu}{v^*},& d&=1-\frac{v^*}{S}=\frac{\mu}{S}>0.
    \label{eq:nominal-constants}
    \end{aligned}
\end{equation}

For the analytical congested diagram in \eqref{mix-fun}, $S=1/(\rho^*h_{\rm mix})$, the derivative definition also applies to the data-based diagram in the experiments. Ignoring higher-order terms yields
\begin{equation}
 \begin{aligned}
 \partial_t\tilde\rho+v^*\partial_x\tilde\rho+\rho^*\partial_x\tilde v&=0,\\
 \partial_t\tilde v-\mu\partial_x\tilde v
 &=-\frac{(S/\rho^*)\tilde\rho+\tilde v}{\tau}.
 \end{aligned}\label{system_rho}
\end{equation}
The linearized physical boundary conditions are
\begin{equation}
 \begin{aligned}
 \tilde\rho(0,t)&=-\frac{\rho^*}{v^*}\tilde v(0,t),\\
 \tilde v(L,t)&=U_b(t)-\frac{v^*}{\rho^*}\tilde\rho(L,t).
 \end{aligned}\label{bound2}
\end{equation}

The characteristic speeds are $\lambda_1=v^*>0$ and $\lambda_2=v^*+\rho^*V'_{\rm mix}(\rho^*)=-\mu<0$. The rest of the paper concerns this heterodirectional congested regime. All stability statements below concern the nominal linearized model with exactly known, constant parameters, ideal actuation and noise-free sensing.

\section{Full-State Feedback Controller}\label{sf-controller}
We parameterize the established ARZ backstepping structure~\cite{yu2019traffic} with the equilibrium and the mixed diagram. The resulting transport speeds and gains depend on the mixed-traffic parameters. We introduce the Riemann coordinate and its scaled form:

\begin{align}
 \tilde w(x,t)&=\frac{S}{\rho^*}\tilde\rho(x,t)+\tilde v(x,t),
 \label{coordinate transformation}\\
 \bar w(x,t)&=e^{x/(\tau v^*)}\tilde w(x,t).
 \label{transformation2}
\end{align}
Here $\tilde w=\tilde v-V'_{\rm mix}(\rho^*)\tilde\rho$ is the traffic states in Riemann coordinates. $\bar w$ is a scaled coordinate for backstepping design.
Define the positive source coefficient
\begin{align}
 c(x)=\frac{1}{\tau}e^{-x/(\tau v^*)}>0.\label{eq:source-coefficient}
\end{align}
The transformed dynamics and boundaries are
\begin{align}
 \begin{aligned}
 \partial_t\bar w+v^*\partial_x\bar w&=0,\\
 \partial_t\tilde v-\mu\partial_x\tilde v&=-c(x)\bar w,
 \end{aligned}\label{transformedsys1}
\end{align}
and the boundary conditions are
\begin{align}
 \begin{aligned}
 \bar w(0,t)&=r_{\mathrm b}\tilde v(0,t),\\
 \tilde v(L,t)&=\frac{1}{r_{\mathrm b}}e^{-L/(\tau v^*)}\bar w(L,t)+U(t),
 \end{aligned}\label{transformedsys4}
\end{align}
where $U(t)=U_b(t)/d$ is the normalized design input. 
The inverse coordinate map recovers the physical states:
\begin{align}
 \begin{aligned}
 \rho(x,t)&=\rho^*+\frac{\rho^*}{S}
 \left[e^{-x/(\tau v^*)}\bar w(x,t)-\tilde v(x,t)\right],\\
 v(x,t)&=v^*+\tilde v(x,t).
 \end{aligned}\label{eq:physical-reconstruction}
\end{align}

Next, we apply the backstepping transformation
\begin{align}
 \begin{aligned}
 \alpha(x,t)&=\bar w(x,t),\\
 \beta(x,t)&=\tilde v(x,t)-\int_0^xK_1(x,\xi)\bar w(\xi,t)d\xi\\
 &\quad-\int_0^xK_2(x,\xi)\tilde v(\xi,t)d\xi,
 \end{aligned}\label{back}
\end{align}

to obtain the target transport cascade
\begin{align}
 \begin{aligned}
    &\partial_t\alpha+v^*\partial_x\alpha=0,\\
 &\partial_t\beta-\mu\partial_x\beta=0,\\
 &\alpha(0,t)=r_{\mathrm b}\beta(0,t),\quad \beta(L,t)=0.
 \end{aligned}\label{targetsys1}
\end{align}

Solving the transport equations by characteristics gives
\begin{align}
 \alpha(x,t)&=\begin{cases}
 \alpha_0(x-v^*t),&t<x/v^*,\\
 r_{\mathrm b}\beta(0,t-x/v^*),&t\ge x/v^*,
 \end{cases}\label{eq:alpha-characteristics}\\
 \beta(x,t)&=\begin{cases}
 \beta_0(x+\mu t),&t<(L-x)/\mu,\\
 0,&t\ge(L-x)/\mu.
 \end{cases}\label{eq:beta-characteristics}
\end{align}
Thus $\beta(x,t)$ clears within $L/\mu$, after which its zero inlet signal clears $\alpha$ within another $L/v^*$. Both target states vanish for $t\ge t_f$, where $t_f=L/v^*+L/\mu$.

To establish exponential decay, use the weighted energy
\begin{align}
 V_{\rm Ly}&=\frac{1}{v^*}\int_0^L(2L-x)\alpha^2dx
 +\frac{2r_{\mathrm b}^2}{\mu}\int_0^L(L+x)\beta^2dx.
 \label{eq:target-energy}
\end{align}
For $\|Z\|_2^2=\int_0^L(\alpha^2+\beta^2)dx$, it satisfies $m_E\|Z\|_2^2\le V_{\rm Ly}\le M_E\|Z\|_2^2$, with $m_E=\min\{L/v^*,2r_{\mathrm b}^2L/\mu\}>0$, $M_E=\max\{2L/v^*,4r_{\mathrm b}^2L/\mu\}$.
Integration by parts and the target boundaries give
\begin{align}
 \begin{aligned}
 \dot V_{\rm Ly}
 &=\frac{2}{v^*}\int_0^L(2L-x)\alpha\partial_t\alpha\,dx\\
 &\quad+\frac{4r_{\mathrm b}^2}{\mu}\int_0^L(L+x)\beta\partial_t\beta\,dx\\
 &=-L\alpha(L,t)^2-\int_0^L(\alpha^2+2r_{\mathrm b}^2\beta^2)dx\\
 &\le-c_0V_{\rm Ly},\qquad c_0=\frac{\min\{v^*,\mu\}}{2L}>0.
 \end{aligned}\label{eq:target-decay}
\end{align}
Using $\mu=S-v^*$, $v^*r_{\mathrm b}=-\mu$ and the positive coefficient $c(x)$ in \eqref{eq:source-coefficient}, differentiation of \eqref{back} and integration by parts yield
\begin{align}
 \begin{aligned}
 \beta_t-\mu\beta_x
 &=[SK_1(x,x)-c(x)]\bar w(x,t)\\
 &\quad+\mu[K_1(x,0)+K_2(x,0)]\tilde v(0,t)\\
 &\quad+\int_0^x[\mu K_{1x}-v^*K_{1\xi}+c(\xi)K_2]
       \bar w(\xi,t)d\xi\\
 &\quad+\mu\int_0^x(K_{2x}+K_{2\xi})\tilde v(\xi,t)d\xi.
 \end{aligned}\label{eq:direct-kernel-check}
\end{align}
The inlet term uses $\bar w(0,t)=r_{\mathrm b}\tilde v(0,t)$. Vanishing diagonal, inlet and integral coefficients gives the following kernel equations and boundary conditions:
\begin{align}
 \begin{aligned}
 \mu\partial_xK_1-v^*\partial_\xi K_1&=-c(\xi)K_2,\\
 \partial_xK_2+\partial_\xi K_2&=0,\\
 K_1(x,x)&=c(x)/S,\\
 K_2(x,0)&=-K_1(x,0).
 \end{aligned}\label{kernel1}
\end{align}
The kernels evolve in the triangular domain $\mathcal T_c=\{(x,\xi):0\le\xi\le x\le L\}$.
The Volterra backstepping transformation \eqref{back} is bounded and boundedly invertible on $(\mathcal L^2(0,L))^2$~\cite{VazquezBacksteppingBoundary2011}, so the plant and target norms are equivalent. The exponential coordinate scaling is also bounded and boundedly invertible on $[0,L]$.

Using the backstepping transformation \eqref{back}, the boundary control law is obtained as
\begin{align}
 U(t)&=\int_0^LK_1(L,\xi)\bar w(\xi,t)d\xi
 +\int_0^LK_2(L,\xi)\tilde v(\xi,t)d\xi\nonumber\\
 &\quad-\frac{1}{r_{\mathrm b}}e^{-L/(\tau v^*)}\bar w(L,t).
 \label{control law}
\end{align}
Therefore, the control law \eqref{control law} stabilizes the transformed mixed-traffic system and drives the traffic states to the equilibrium point at finite time $t_f$. Let $X = (\bar w,\tilde v)$, we have the following theorem.

\begin{thm}\label{thm1}
For the nominal linearized system \eqref{transformedsys1}--\eqref{transformedsys4}, with $v^*>0$, $\mu>0$, exact known constant parameters and ideal actuation, let $X_0=(\bar w_0,\tilde v_0)\in (\mathcal L^2(0,L))^2$. The state-feedback law \eqref{control law} gives a unique mild solution and constants $M_s,\eta_s>0$ such that
\begin{align}
 \|X(t)\|_2\le M_s e^{-\eta_st}\|X_0\|_2, X(t)=0\text{ for }t\ge t_f.
 \label{eq:sf-stability-bound}
\end{align}
The constants depend only on the fixed nominal parameters and kernel bounds.
\end{thm}
\begin{proof}
The boundary control law~\eqref{control law} enforces $\beta(L,t)=0$ in the target system. Its characteristic solution is unique and clears both target states by $t_f$. The system~\eqref{eq:target-decay} gives exponential energy decay. Norm equivalence of the direct and inverse transformations gives \eqref{eq:sf-stability-bound}. The physical coordinate map \eqref{eq:physical-reconstruction} is bounded and invertible for the fixed parameters, so the same convergence holds for the physical traffic density and speed.
\end{proof}

\section{Output Feedback Controller}\label{of-controller}
\subsection{Collocated State Observer}\label{sec:observer}
Full-state sensing is a strong assumption for freeway control. Loop detectors, cameras and connected-vehicle reports may provide rich data, but full in-domain density and speed are difficult to obtain. We therefore design a state observer at the same boundary where the control acts. The ideal collocated measurement is
\begin{align}
 y_1(t)=\bar w(L,t).\label{eq:boundary-output}
\end{align}
The theoretical analysis assumes synchronized, noise-free boundary sensing.The effect of sensing errors is examined in the sensor-noise experiments in Section~\ref{sec:sensor-noise}.
The observer dynamics are
\begin{align}
 \begin{aligned}
 \partial_t\hat w+v^*\partial_x\hat w
 &=p_1(x)[y_1(t)-\hat w(L,t)],\\
 \partial_t\hat v-\mu\partial_x\hat v
 &=-c(x)\hat w+p_2(x)[y_1(t)-\hat w(L,t)],
 \end{aligned}\label{obs1}
\end{align}
with boundary conditions
\begin{align}
 \begin{aligned}
 \hat w(0,t)&=r_{\mathrm b}\hat v(0,t),\\
 \hat v(L,t)&=\frac{1}{r_{\mathrm b}}e^{-L/(\tau v^*)}y_1(t)+U(t).
 \end{aligned}\label{obs2}
\end{align}
Here $\hat w$ and $\hat v$ estimate $\bar w$ and $\tilde v$. Subtracting observer from plant and defining $\breve w=\bar w-\hat w$ and $\breve v=\tilde v-\hat v$, then we have
\begin{align}
 \begin{aligned}
 \partial_t\breve w+v^*\partial_x\breve w&=-p_1(x)\breve w(L,t),\\
 \partial_t\breve v-\mu\partial_x\breve v&=-c(x)\breve w-p_2(x)\breve w(L,t),\\
 \breve w(0,t)&=r_{\mathrm b}\breve v(0,t),\qquad \breve v(L,t)=0.
 \end{aligned}\label{eq:observer-error-system}
\end{align}
Denoting error-target variables as $(\gamma,\delta)$ and using another Volterra transformation
\begin{align}
 \begin{aligned}
 \breve w(x,t)&=\gamma(x,t)-\int_x^LM_1(x,\xi)\gamma(\xi,t)d\xi,\\
 \breve v(x,t)&=\delta(x,t)-\int_x^LM_2(x,\xi)\gamma(\xi,t)d\xi.
 \end{aligned}\label{obtrans1}
\end{align}
The target satisfies \eqref{targetsys1} with $(\alpha,\beta)$ replaced by $(\gamma,\delta)$.

Differentiating \eqref{obtrans1} and integrating by parts gives
\begin{align}
 \begin{aligned}
 \breve w_t+v^*\breve w_x+p_1\gamma(L,t)
 &=[v^*M_1(x,L)+p_1]\gamma(L,t)\\
 &\quad-v^*\int_x^L(M_{1x}+M_{1\xi})\gamma(\xi,t)d\xi,
 \end{aligned}\label{eq:observer-first-residual}
\end{align}
and
\begin{align}
 \begin{aligned}
 &\breve v_t-\mu\breve v_x+c(x)\breve w+p_2\gamma(L,t)\\
 &\quad=[v^*M_2(x,L)+p_2]\gamma(L,t)\\
 &\qquad-[SM_2(x,x)-c(x)]\gamma(x,t)\\
 &\qquad+\int_x^L[\mu M_{2x}-v^*M_{2\xi}-c(x)M_1]
             \gamma(\xi,t)d\xi.
 \end{aligned}\label{eq:observer-second-residual}
\end{align}
The condition $M_1(0,\xi)=r_{\mathrm b}M_2(0,\xi)$ transfers $\gamma(0,t)=r_{\mathrm b}\delta(0,t)$ to $\breve w(0,t)=r_{\mathrm b}\breve v(0,t)$. At the outlet, $\breve w(L,t)=\gamma(L,t)$ and $\breve v(L,t)=0$. Vanishing coefficients gives the kernel equations on the domain $\mathcal T_o=\{(x,\xi):0\le x\le\xi\le L\}$:

\begin{align}
 \begin{aligned}
 \partial_xM_1+\partial_\xi M_1&=0,\\
 \mu\partial_xM_2-v^*\partial_\xi M_2&=c(x)M_1,\\
 M_1(0,\xi)&=r_{\mathrm b}M_2(0,\xi),\\
 M_2(x,x)&=c(x)/S.
 \end{aligned}\label{obsker1}
\end{align}
Choose the observer gains as
\begin{align}
 p_1(x)&=-v^*M_1(x,L),&p_2(x)&=-v^*M_2(x,L).
 \label{obsgain1}
\end{align}
Let $\widehat{X} = (\hat w,\hat v)$, then we have the following theorem.
\begin{thm}\label{thm:observer}
Under the nominal parameter assumptions of Theorem~\ref{thm1}, let $X_0,\widehat X_0\in (\mathcal L^2(0,L))^2$. With exact output \eqref{eq:boundary-output}, the same known applied input in plant and observer, and gains \eqref{obsgain1}, the error $X_{\mathrm e}=X-\widehat X=(\breve w,\breve v)$ satisfies
\begin{align}
 \|X_{\mathrm e}(t)\|_2\le M_e e^{-\eta_et}\|X_{\mathrm e}(0)\|_2,
 \qquad X_{\mathrm e}(t)=0 \text{ for } t\ge t_f,
 \label{eq:observer-stability-bound}
\end{align}
where $M_e,\eta_e>0$ depend only on the nominal parameters and kernels.
\end{thm}
\begin{proof}
With the gains \eqref{obsgain1}, the observer-error system is mapped by \eqref{obtrans1} to the same target cascade system used in the state-feedback design. The target variables therefore vanish after $t_f$. Bounded invertibility of the observer-error Volterra transformation on $L_2(0,L)$ implies $\breve{w}(x,t)=0$ and $\breve{v}(x,t)=0$ for all $x\in[0,L]$ once $t\ge t_f$.
\end{proof}

The state-feedback law is implemented using the estimated states to form the output-feedback law as:
\begin{align}
 \begin{aligned}
 U(t)&=\int_0^LK_1(L,\xi)\hat w(\xi,t)d\xi
 +\int_0^LK_2(L,\xi)\hat v(\xi,t)d\xi\\
 &\quad-\frac{1}{r_{\mathrm b}}e^{-L/(\tau v^*)}\hat w(L,t).
 \end{aligned}\label{output control law}
\end{align}
The following theorem establishes the stability of the output-feedback closed-loop system.
{
\begin{thm}\label{thm:output-feedback}\label{thm:output}
Under the nominal assumptions of Theorem~\ref{thm:observer}, the output-feedback law \eqref{output control law} gives a unique mild solution for $\Xi=(X,X_{\mathrm e})$ and constants $M_o,\eta_o>0$ such that
\begin{align}
 \begin{aligned}
 \|\Xi(t)\|_2&\le M_o e^{-\eta_ot}\|\Xi(0)\|_2,\\
 X(t)&=X_{\mathrm e}(t)=0\text{ for }t\ge2t_f,
 \end{aligned}\label{eq:of-stability-bound}
\end{align}
where $\|\Xi\|_2=(\|X\|_2^2+\|X_{\mathrm e}\|_2^2)^{1/2}$.
\end{thm}
\begin{proof}
Let $(\alpha,\beta)$ be the plant targets and $(\gamma,\delta)$ the observer-error targets. Define
\begin{align}
 \begin{aligned}
 F(X_{\mathrm e})&=\int_0^L[K_1(L,\xi)\breve w(\xi)
                  +K_2(L,\xi)\breve v(\xi)]d\xi,\\
 b_L&=\frac{1}{r_{\mathrm b}}e^{-L/(\tau v^*)}.
 \end{aligned}\label{eq:of-error-functional}
\end{align}
Substitution of \eqref{output control law} into the plant boundary gives
\begin{align}
 \beta(L,t)=d(t)=b_L\breve w(L,t)-F(X_{\mathrm e}(t)).
 \label{eq:of-target-boundary}
\end{align}
The observer error is independent of the shared applied input. Its target has $\delta(L,t)=0$, and \eqref{obtrans1} gives $\breve w(L,t)=\gamma(L,t)$.

Let $V_p$ and $V_e$ be \eqref{eq:target-energy} on the plant and error targets. Integration by parts gives
\begin{align}
 \begin{aligned}
 \dot V_p&\le-c_0V_p-L\alpha(L,t)^2+4L r_{\mathrm b}^2d(t)^2,\\
 \dot V_e&\le-c_0V_e-L\gamma(L,t)^2.
 \end{aligned}\label{eq:of-energy-estimates}
\end{align}
Set $k_F^2=\int_0^L[K_1(L,\xi)^2+K_2(L,\xi)^2]d\xi$. Boundedness of the observer transformation and positivity of $V_e$ give $C_e>0$ with $\|X_{\mathrm e}\|_2^2\le C_eV_e$. Consequently,
\begin{align}
 d^2&\le A\gamma(L,t)^2+B V_e,
 \label{eq:of-boundary-estimate}
\end{align}
where $A=2b_L^2$, $B=2k_F^2C_e$. All constants depend only on the fixed parameters and kernel norms. Choose $\kappa\ge\max\{1,4r_{\mathrm b}^2A,8L r_{\mathrm b}^2B/c_0\}$, and then define the augmented energy
\begin{align}
 W&=V_p+\kappa V_e.
 \label{eq:of-augmented-energy}
\end{align}
Combining the estimates gives
\begin{align}
 \begin{aligned}
 \dot W&\le-c_0V_p-(\kappa c_0-4L r_{\mathrm b}^2B)V_e\\
 &\quad-(\kappa L-4L r_{\mathrm b}^2A)\gamma(L,t)^2\\
 &\le-(c_0/2)W.
 \end{aligned}\label{eq:of-composite-decay}
\end{align}
Norm equivalence of both transformations establishes the augmented exponential estimate, with a possible norm decay rate $\eta_o = c_0/4$. By Theorem~\ref{thm:observer}, $X_{\mathrm e}$ and $h$ vanish for $t\ge t_f$. The plant then clears in another $t_f$, establishing extinction by $2t_f$.
\end{proof}
}

\section{Experiments}\label{simulation}
\subsection{Simulation Configuration}\label{sec:configuration}
We simulate the nominal linearized ARZ system on a $L=500$ m segment for $400$ s. The equilibrium density is $\rho^*=45$ veh/km/lane, and each scenario uses its own diagram-consistent $v^*$, as listed in Table~\ref{equilibrium state}. The adaptation time is $\tau=60$ s, following an ARZ numerical scenario in~\cite{yu2019traffic}. The initial condition is a sinusoidal perturbation around the equilibrium state to mimic the stop-and-go congestion, as given in \eqref{initial}
\begin{align}
\begin{aligned}
\rho(x,0)&=\rho^*\left[1+0.1\sin(3\pi x/L)\right],\\
v(x,0)&=v^*\left[1-0.1\sin(3\pi x/L)\right].
\end{aligned}\label{initial}
\end{align}
Each tested scenario are under the same intital condition and the inlet flow $q^* = \rho^*v^*$. Observer perturbations start at zero. Observer interior perturbations start at zero. For the numcerical implementation, the spatial grid consists of 202 nodes, including 200 interior nodes and two boundary nodes, yielding 201 intervals with $\Delta x=500/201=2.49$ m. Second-order upwind differences are used in the simulation. The time step is $\Delta t=0.1$ s, which satisfies the CFL condition for the transport speeds. The simulations are run on an Interl Core i9-12900K CPU(5.20 GHZ), an NVIDIA GeForce RTX 3090 GPU, 2.0TB of storage, and 64 GB of memory. All code and results necessary for reproducibility are available at \url{https://github.com/curryzyang/Stabilization4MixedTraffic.git}.

{ We adapt the proportional-integral(PI) controller~\cite{ZhangPI2019} as a benchmark for comparison. It is a one-sided benchmark adaptation, not a replication of that work's two-actuator setting or stability theorem. Specifically, we design the following PI controller for the downstream boundary speed regulation:
\begin{align}
\begin{aligned}
U_{\rm PI}(t)&=k_p(v(0,t)-v^*)+k_i \int_0^t (v(0,s)-v^*)ds,
\end{aligned}\label{eq:pi-law}
\end{align}
where fixed gains are $k_p=0.1$ and $k_i=-0.002 \mathrm{s}^{-1}$. All physical state plots and traffic indices use \eqref{eq:physical-reconstruction}. In the following, we denote the full-state feedback controller as SF, the output-feedback controller as OF, and the PI controller as PI for brevity.
}
\begin{table}[tbp]
\centering\small
\caption{Equilibria for the nominal penetration scenarios}\label{equilibrium state}
\setlength{\tabcolsep}{3pt}
\begin{tabular}{lrrrrrr}\hline
$P$ & 0 & 0.2 & 0.4 & 0.6 & 0.8 & 1\\\hline
$\rho^*$ (veh/km/lane) & 45 & 45 & 45 & 45 & 45 & 45\\
$v^*$ (km/h) & 36.53 & 39.71 & 43.49 & 48.07 & 53.73 & 60.89\\
\hline\end{tabular}\end{table}

\subsection{Closed-Loop Results}\label{sec:closed-loop}
{Figures~\ref{clsf-06}--\ref{fig:pi-state} show traffic density and speed at $P=0.6$. The OF figure uses its own plant states, independently of the SF states and observer estimates. SF and OF damp the initial waves, with a longer observer-related transient for OF. PI reduces the initial disturbance but leaves visible residual waves during the simulation period.
\begin{figure}[tbp]\centering
\subfloat[Density]{\includegraphics[width=.48\linewidth]{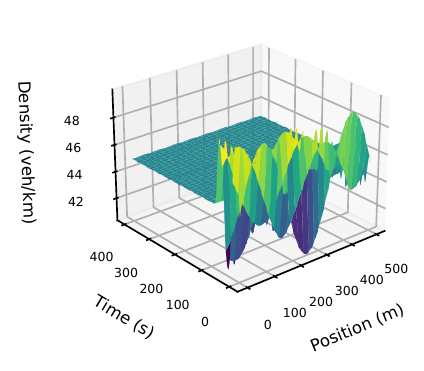}}
\subfloat[Speed]{\includegraphics[width=.48\linewidth]{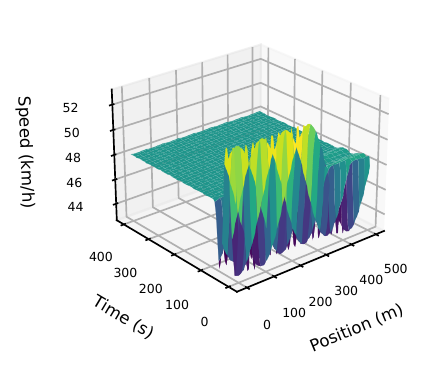}}
\caption{Physical SF states at $P=0.6$.}\label{clsf-06}\end{figure}
\begin{figure}[tbp]\centering
\subfloat[Density]{\includegraphics[width=.48\linewidth]{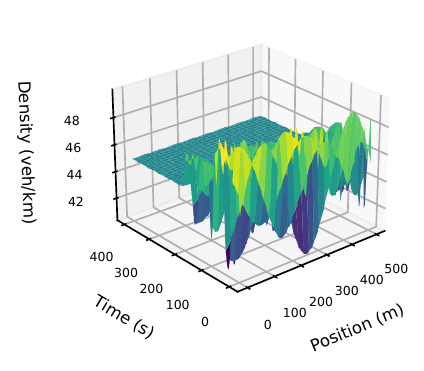}}
\subfloat[Speed]{\includegraphics[width=.48\linewidth]{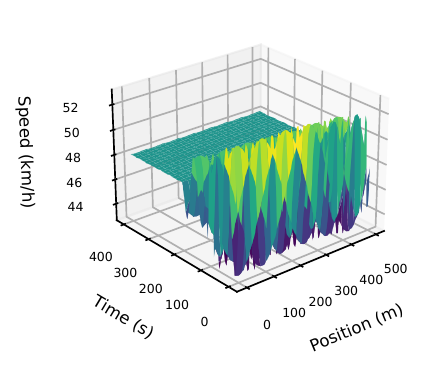}}
\caption{Physical OF plant states at $P=0.6$.}\label{clof-06}\end{figure}
\begin{figure}[tbp]\centering
\subfloat[Density]{\includegraphics[width=.48\linewidth]{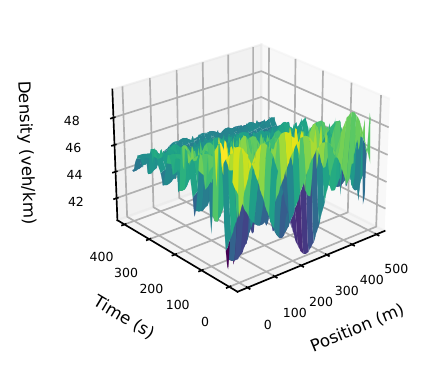}}
\subfloat[Speed]{\includegraphics[width=.48\linewidth]{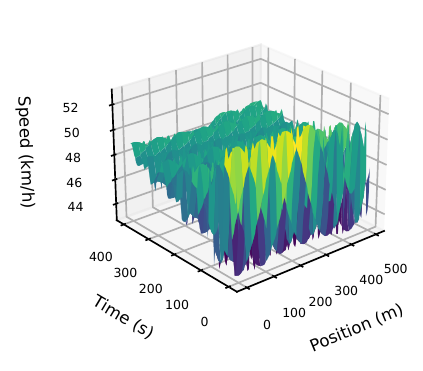}}
\caption{Physical states under the PI controller at $P=0.6$.}\label{fig:pi-state}\end{figure}
}
\subsection{Convergence Speed}\label{sec:convergence}
{To compare the convergence speed for different mehtods under different penetration rates, we measure the settling time of each method. Measured settling time is based on the relative physical-state norm
\begin{align}
E_{\text{plant}}(t)=\left(\int_0^L\left[\left(\frac{\rho-\rho^*}{\rho^*}\right)^2+
\left(\frac{v-v^*}{v^*}\right)^2\right]dx\right)^{1/2}.
\label{eq:settling-norm}
\end{align}
It is the first saved time after the last threshold exceedance such that $E_{\text{plant}}(t)\le10^{-4}\,\mathrm{m}^{1/2}$ over the $400$ s horizon. A trajectory that still exceeds the threshold at $400$ s is reported as $>400$. Relative traffic state terms are dimensionless, so spatial integration gives $E_{\text{plant}}$ the unit $\mathrm{m}^{1/2}$. Table~\ref{convergence} distinguishes analytical settling time bounds $t_f=L/v^*+L/\mu$ and $2t_f$ from the measured discrete settling times.
\begin{table}[tbp]
\centering\small
\caption{Analytical settling time and measured settling time (s)}\label{convergence}
\setlength{\tabcolsep}{2.5pt}
\begin{tabular}{lrrrrrr}\hline
$P$ & 0 & 0.2 & 0.4 & 0.6 & 0.8 & 1\\\hline
SF analytical & 156.4 & 143.9 & 131.4 & 118.9 & 106.4 & 93.8\\
SF measured & 180.0 & 163.1 & 145.9 & 126.6 & 112.5 & 99.1\\
OF analytical & 312.8 & 287.8 & 262.8 & 237.7 & 212.7 & 187.7\\
OF measured & 320.6 & 295.2 & 269.7 & 244.2 & 218.7 & 193.1\\
PI measured & $>400$ & $>400$ & $>400$ & $>400$ & $>400$ & $>400$\\
\hline\end{tabular}
\end{table}
Increasing the nominal penetration rate shortens the characteristic travel times and also reduces the measured settling times of the SF and OF controllers in these scenarios. By contrast, the PI controller remains above the $10^{-4}$ threshold at every penetration rate throughout the $400$ s horizon, so it does not satisfy the measured-settling criterion in this setting. Figure~\ref{fig:pi-benchmark} compares the physical state norms using the same $10^{-4}$ threshold for all controllers at $P=0.6$. The figure shows that the PI controller converges more slowly than the SF and OF controllers, which further supports the improved convergence performance of the proposed control method.
\begin{figure}[tbp]\centering
\includegraphics[width=\linewidth]{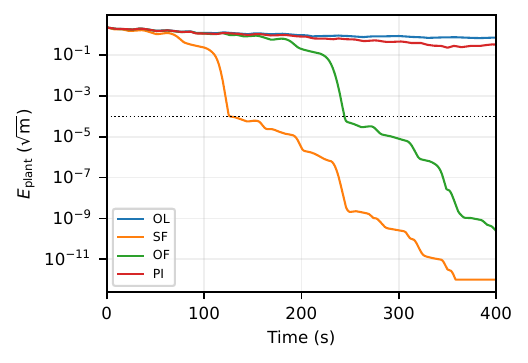}
\caption{Physical state norms at $P=0.6$.}\label{fig:pi-benchmark}\end{figure}
}

\subsection{Control Effort and Traffic Performance}\label{sec:metrics}
{Figure~\ref{control effort} compares nominal physical boundary inputs $U_b$. At $P=0.6$, PI has a peak input magnitude of $0.078$ m/s, compared with $0.427$ m/s for SF and $0.248$ m/s for OF. Its smaller input is accompanied by slower regulation and a residual state error.
\begin{figure}[tbp]\centering
\includegraphics[width=\linewidth]{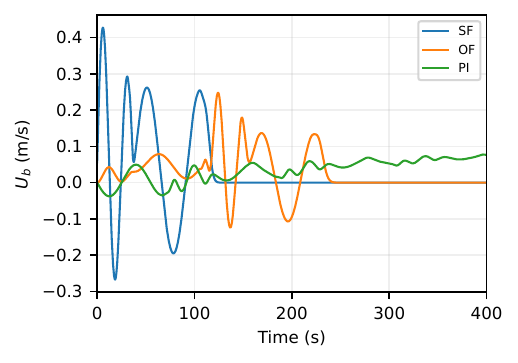}
\caption{Physical boundary inputs $U_b$ under SF, OF and PI.}\label{control effort}\end{figure} 
Next, we use several traffic performance indices to compare the controllers, including fuel consumption, total travel time (TTT), driver discomfort, and total delay , as introduced in~\cite{treiber2013traffic}. With vehicle acceleration $a_{\rm veh}=v_t+vv_x$, the fuel-rate proxy is
\begin{align}
f(v,a_{\rm veh})=\max\{0,b_0+b_1v+b_2va_{\rm veh}+b_3v^3\}.
\label{eq:fuel-proxy}
\end{align}
The SI coefficients $b_0=0.025$, $b_1=24.5\times10^{-6}$, $b_2=125\times10^{-6}$ and $b_3=32.5\times10^{-9}$ yield a fuel rate in L/s per vehicle~\cite[p.~485]{treiber2013traffic}. We compute
\begin{align}
\begin{aligned}
J_{\rm fuel}&=\int_0^T\!\int_0^L\rho f(v,a_{\rm veh})\,dx\,dt,\\
D&=10^{-3}\int_0^T\!\int_0^L\rho v\,dx\,dt,\quad I_{\rm fuel}=J_{\rm fuel}/D,\\
J_{\rm TTT}&=\int_0^T\!\int_0^L\rho\,dx\,dt,\\
J_{\rm delay}&=J_{\rm TTT}-1000D/v_f,\\
J_{\rm discom}&=\int_0^T\!\int_0^L\rho\left[a_{\rm veh}^2+
(\partial_t a_{\rm veh})^2\right]dx\,dt.
\end{aligned}\label{eq:performance-metrics}
\end{align}
Here, $D$ is vehicle-kilometres, $I_{\rm fuel}$ is the fuel consumption for pervehicel perkilometers with units of L/vehicle-km, $J_{\rm TTT}$ and $J_{\rm delay}$ are veh$\cdot$s, and $J_{\rm discom}$ has units $\mathrm{veh}\cdot\mathrm{m}^2/\mathrm{s}^3$. The term $\partial_t a_{\rm veh}$ denotes a partial derivative with respect to time. The results for OL, SF, OF, and PI are shown in Table~\ref{tab:performance_output}. It contains each penetration's absolute OL baseline and the percentage changes of SF, OF, and PI relative to it. For any cost $J$,
\begin{align}
\Delta J(P)=100\frac{J_{\rm mode}(P)-J_{\rm OL}(P)}{J_{\rm OL}(P)}.
\label{eq:matched-change}
\end{align}

\begin{table*}[tbp]
\centering\small
\caption{Traffic performance under different control strategies: absolute OL baselines and closed-loop changes}\label{tab:performance_output}
\setlength{\tabcolsep}{2.5pt}
\begin{tabular}{c l rrrrr}
\hline
$P$ & Mode & Fuel(L) & Fuel(L/veh-km) & Discomfort Value & TTT(veh s) & Total Delay(veh s) \\
\hline
0 & OL (abs.) & 232.34 & 2.55 & 183.08 & 9191.21 & 6455.57 \\
0 & SF (\%) & -1.71 & -1.94 & -70.87 & -1.73 & -2.57 \\
0 & OF (\%) & -1.28 & -1.47 & -11.80 & -1.30 & -1.93 \\
0 & PI (\%) & -0.63 & -0.81 & -0.27 & -0.65 & -1.00 \\
\hline
0.2 & OL (abs.) & 232.63 & 2.35 & 255.93 & 9191.21 & 6217.39 \\
0.2 & SF (\%) & -1.73 & -1.95 & -73.43 & -1.76 & -2.70 \\
0.2 & OF (\%) & -1.32 & -1.50 & -14.47 & -1.34 & -2.07 \\
0.2 & PI (\%) & -0.67 & -0.83 & -0.35 & -0.69 & -1.10 \\
\hline
0.4 & OL (abs.) & 232.99 & 2.15 & 368.43 & 9191.21 & 5933.82 \\
0.4 & SF (\%) & -1.75 & -1.95 & -75.90 & -1.78 & -2.87 \\
0.4 & OF (\%) & -1.36 & -1.52 & -17.21 & -1.39 & -2.24 \\
0.4 & PI (\%) & -0.70 & -0.87 & -0.34 & -0.73 & -1.23 \\
\hline
0.6 & OL (abs.) & 233.45 & 1.95 & 552.86 & 9191.21 & 5590.52 \\
0.6 & SF (\%) & -1.77 & -1.96 & -78.41 & -1.81 & -3.09 \\
0.6 & OF (\%) & -1.40 & -1.55 & -20.54 & -1.43 & -2.45 \\
0.6 & PI (\%) & -0.75 & -0.91 & -0.30 & -0.78 & -1.39 \\
\hline
0.8 & OL (abs.) & 234.09 & 1.75 & 873.60 & 9191.21 & 5167.33 \\
0.8 & SF (\%) & -1.79 & -1.98 & -80.97 & -1.83 & -3.41 \\
0.8 & OF (\%) & -1.45 & -1.60 & -24.55 & -1.48 & -2.76 \\
0.8 & PI (\%) & -0.81 & -0.96 & -0.24 & -0.85 & -1.64 \\
\hline
1 & OL (abs.) & 234.97 & 1.55 & 1465.32 & 9191.21 & 4630.15 \\
1 & SF (\%) & -1.81 & -1.98 & -83.47 & -1.86 & -3.86 \\
1 & OF (\%) & -1.49 & -1.62 & -28.94 & -1.54 & -3.18 \\
1 & PI (\%) & -0.87 & -1.02 & +0.01 & -0.93 & -2.00 \\
\hline
\end{tabular}
\end{table*}
Across penetration rates, Table~\ref{tab:performance_output} shows
competing trends among the traffic performance indices. As $P$ increases, total delay and fuel consumption per vehicle-kilometre decrease, whereas total fuel consumption and the discomfort value increase. For example, under SF, increasing $P$ from $0$ to $1$ reduces fuel consumption per vehicle-kilometre from $2.50$ to $1.52$ L/vehicle-km, while total fuel consumption increases from $228.37$ to $230.73$ L. The reduction in distance-normalized fuel consumption indicates improved fuel efficiency. However, because 
vehicle-distance traveled over the fixed simulation horizon increses, the aggregate fuel consumption also increases. Thus, higher penetration improves delay and fuel efficiency in these scenarios, accompanied by increased total fuel consumption and discomfort.

At a fixed penetration rate, the mathced-baseline comparisons isolate the benefits of boundary control. Both SF and OF reduce total fuel consumption, fuel consumption per vehicle-kilometre, discomfort, total travel time and total delay relative to the corresponding OL case. At $P=1$, SF and OF reduce delay by $3.86\%$ and $3.18\%$,
distance-normalized fuel consumption by $1.98\%$ and $1.62\%$, and discomfort by $83.47\%$ and $28.94\%$, respectively. Both controllers also achieve larger improvements than PI controller across the reported traffic performance indices. These comparisons demonstrate the benefits of the proposed boundary-feedback schemes
under the same penetration rate and operating conditions.

These results suggest that boundary controllers and CAV penetration levels should be selected jointly to balance improvements in delay and fuel efficiency against total fuel consumption and driving discomfort.
}
\subsection{Endpoint-Parameter Sensitivity}\label{sec:endpoint-sensitivity}
{In this section, we examine the sensitivity of the closed-loop system to the endpoint parameters. Two separate analyses keep $P=1$. The CAV gap takes values of $,2,$ and $3$ m with $h_1=0.9$ s, while the headway takes values of $0.6, 0.9,$ and $1.1$s with $s_g^1=2$ m. In each scenario, we recompute the equilibrium and SF/OF kernels, and compares the controllers with the corresponding OL case using the same initial conditions. Tables~\ref{tab:gap-sensitivity} and~\ref{tab:headway-sensitivity} report capacity, equilibrium speed, settling time and delay reduction $100(J_{\rm OL}-J_{\rm mode})/J_{\rm OL}$.
\begin{table}[tbp]
\centering\scriptsize
\caption{CAV gap sensitivity at $P=1$, with $h_1=0.9$ s}\label{tab:gap-sensitivity}
\setlength{\tabcolsep}{2.5pt}
\begin{tabular}{c r r cc cc}
\hline
$s_g^1$ (m) & $Q_{\max}$ & $v^*$ & \multicolumn{2}{c}{Settling time (s)} & \multicolumn{2}{c}{Delay reduction (\%)} \\
 & (veh/h/lane) & (km/h) & SF & OF & SF & OF \\
\hline
1 & 3333.3 & 64.89 & 108.6 & 211.6 & 4.09 & 3.27 \\
2 & 3243.2 & 60.89 & 99.1 & 193.1 & 3.86 & 3.18 \\
3 & 3157.9 & 56.89 & 92.5 & 180.6 & 3.65 & 3.08 \\
\hline
\end{tabular}
\end{table}
Increasing the standstill gap lowers model capacity and equilibrium speed, while numerical settling shortens because the upstream characteristic speed increases. 
\begin{table}[tbp]
\centering\scriptsize
\caption{CAV headway sensitivity at $P=1$, with $s_g^1=2$ m}\label{tab:headway-sensitivity}
\setlength{\tabcolsep}{2.5pt}
\begin{tabular}{c r r cc cc}
\hline
$h_1$ (s) & $Q_{\max}$ & $v^*$ & \multicolumn{2}{c}{Settling time (s)} & \multicolumn{2}{c}{Delay reduction (\%)} \\
 & (veh/h/lane) & (km/h) & SF & OF & SF & OF \\
\hline
0.6 & 4444.4 & 91.33 & 65.9 & 128.9 & 7.98 & 6.93 \\
0.9 & 3243.2 & 60.89 & 99.1 & 193.1 & 3.86 & 3.18 \\
1.1 & 2748.1 & 49.82 & 121.7 & 235.7 & 3.19 & 2.54 \\
\hline
\end{tabular}
\end{table}
Increasing headway likewise lowers capacity and equilibrium speed, lengthens the characteristic travel times and measured settling, and reduces the delay benefit.
}
\subsection{VSL Magnitude and Rate Limits}\label{sec:vsl-constraints}
{In practice, the actuation of VSL systems is subject to physical limitations. In this section, we impose execution constraints directly on the controller's design input $U$, with the physical command recovered through the same coordinate map. For SF or OF, $U_{\rm nom}$ is the nominal input obtained by~\eqref{control law} and \eqref{output control law}. The applied input to the system satisfy
\begin{align}
\begin{aligned}
U_{\min}\le U_{\rm app}\le U_{\max},\quad
|\dot U_{\rm app}|\le r_U,
\end{aligned}\label{eq:vsl-input-map}
\end{align}
where $U_{\min}$, $U_{\max}$, and $r_U$ are the minimum and maximum allowed values for the applied input and its rate of change, respectively. Plant and observer receive exactly the same applied input. Both scenarios use $P=0.6$, $U_{\min}=-0.3$ m/s and $U_{\max}=0.3$ m/s. Tight magnitude applies $U_{\rm app}=\operatorname{sat}_{[U_{\min},U_{\max}]}(U_{\rm nom})$. Tight magnitude plus rate additionally uses $r_U=0.03$ m/s$^2$ and $U_{\rm app}(0)=0$. The nominal SF/OF input ranges and rates exceed them, ensuring actual constraint activation. Let $U_{\mathrm{sat}}(t)=\operatorname{sat}_{[U_{\min},U_{\max}]}(U_{\rm nom}(t))$. Between switching instants, the applied input evolves according to
\begin{align}
\dot U_{\rm app}=
\begin{cases}
r_U\operatorname{sgn}(U_{\mathrm{sat}}-U_{\rm app}),&U_{\mathrm{sat}}\ne U_{\rm app},\\
\operatorname{sat}_{[-r_U,r_U]}(\dot U_{\mathrm{sat}}),&U_{\mathrm{sat}}=U_{\rm app}.
\end{cases}\label{eq:vsl-actuator}
\end{align}
Once the applied input reaches the target, it follows the target as long as the target's rate of change remains within the prescribed bound. Otherwise, it approaches the target at the maximum allowed rate, including during the initial transient.
\begin{table*}[tbp]
\centering\small
\caption{Strict design-input limits at $P=0.6$}\label{tab:actuator}
\setlength{\tabcolsep}{2.5pt}
\begin{tabular}{llrrrrr}\hline
Scenario & Mode & Settling (s) & Delay reduction (\%) &\shortstack{Fuel/veh-km\\(L/(veh km))} & \shortstack{Magnitude\\activation (\%)} & \shortstack{Rate\\activation (\%)}\\\hline
Tight magnitude & SF & 350.3 & 2.43 & 1.9149 & 32.00 & --\\
Tight magnitude & OF & 353.4 & 2.27 & 1.9172 & 15.30 & --\\
Tight magnitude + rate & SF & 394.0 & 2.52 & 1.9141 & 34.98 & 43.61\\
Tight magnitude + rate & OF & 392.9 & 2.35 & 1.9163 & 15.53 & 28.67\\
\hline\end{tabular}
\end{table*}
Table~\ref{tab:actuator} reports settling and traffic performance, with magnitude activation denoting the time fraction when the nominal input exceeds the amplitude bounds and rate activation the fraction spent ramping at the maximum allowed slope, both over the full $400$ s horizon. The active limits prolong SF/OF settling from $126.6/244.2$ s to $350.3/353.4$ s (magnitude) and $394.0/392.9$ s (magnitude plus rate), while all four cases reach the common threshold within $400$ s.
\begin{figure}[tbp]\centering
\includegraphics[width=\linewidth]{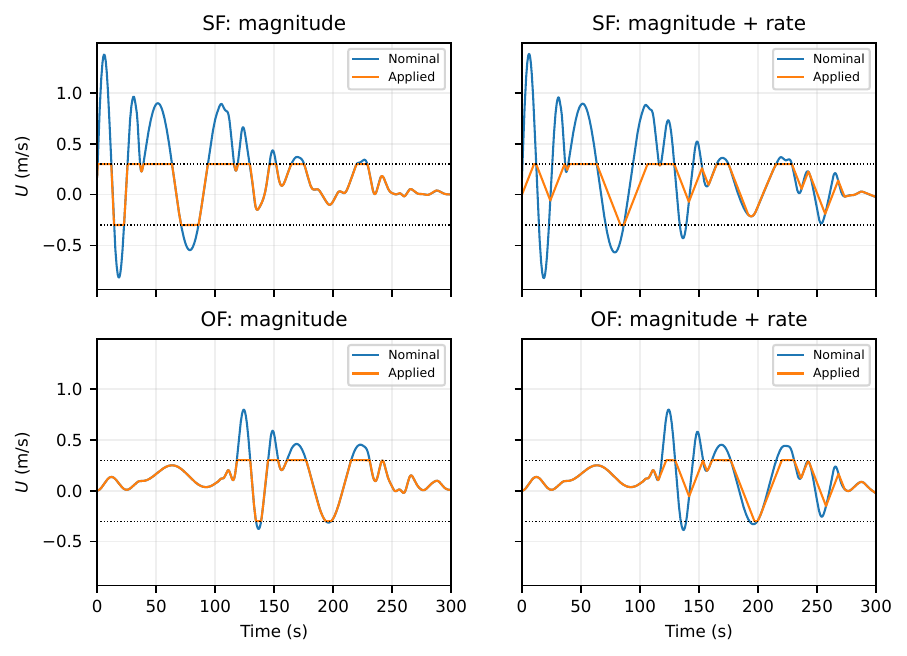}
\caption{Applied control inputs $U_{\rm app}$ under tight magnitude and magnitude-plus-rate constraints at $P=0.6$.}\label{fig:actuator-input}\label{fig:vsl-constraints}\end{figure}
Figure~\ref{fig:actuator-input} shows the applied design inputs $U_{\rm app}$ under the two constraint scenarios. The SF/OF nominal inputs exceed the bounds, and the applied input follows the target at the maximum allowed rate until it reaches the target. The magnitude-plus-rate scenario has a longer ramp phase than the magnitude-only scenario, which is consistent with its longer settling time.
Figure~\ref{fig:vsl-norm} shows the corresponding physical perturbation norms. Both SF and OF converge to the threshold, with a longer transient than their nominal case.
\begin{figure}[tbp]\centering
\subfloat[SF]{\includegraphics[width=.48\linewidth]{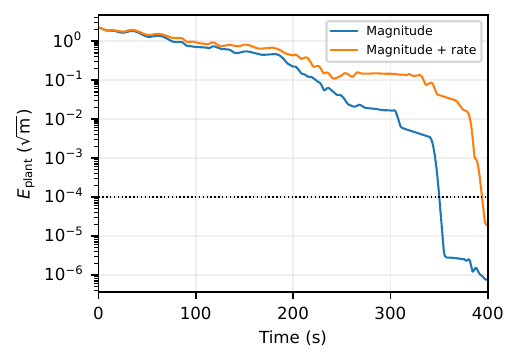}}
\subfloat[OF]{\includegraphics[width=.48\linewidth]{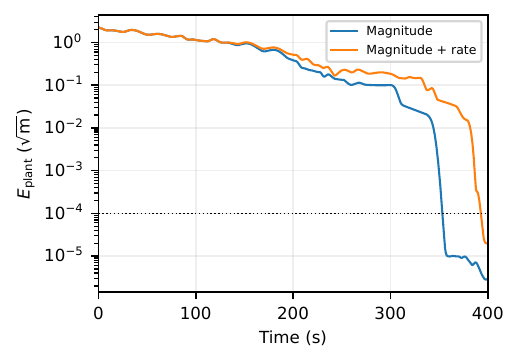}}
\caption{Physical state norms under the two tight input constraints.}\label{fig:vsl-norm}\end{figure}

The limited-input trajectories provide finite-horizon numerical evidence. For state feedback, the target boundary becomes $\beta(L,t)=U_{\rm app}-U_{\rm nom}$, leading to a persistent mismatch that invalidates nominal finite-time extinction. A constrained stability analysis would need boundary-input estimates and feasibility conditions, with observer coupling included for OF. Saturated hyperbolic-PDE feedback~\cite{Shreim2022} and finite-dimensional amplitude/rate-limited control~\cite{GomesDaSilva2003} offer methodological starting points, but a complete treatment is beyond this paper's scope. The numerical results show that the backstepping design can tolerate realistic VSL actuation limits.
}
\subsection{Physical Boundary Sensor Noise}\label{sec:sensor-noise}
\begin{figure}[tbp]\centering
\subfloat[Plant regulation]{\includegraphics[width=.48\linewidth]{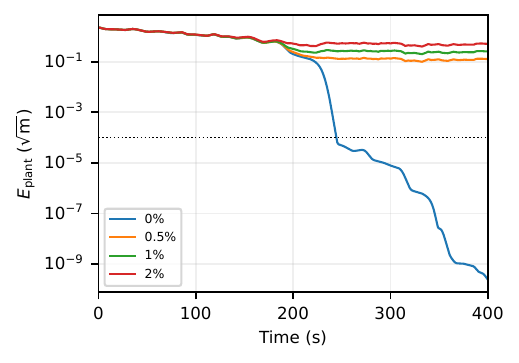}}
\subfloat[Observer error]{\includegraphics[width=.48\linewidth]{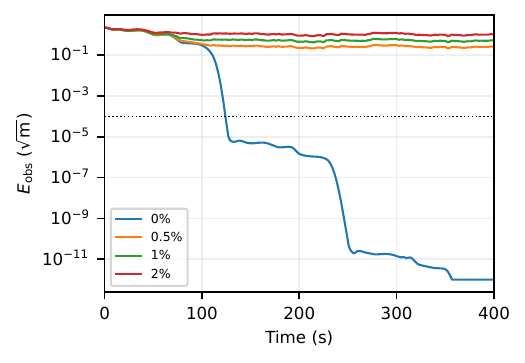}}
\caption{Physical sensor-error norms at $P=0.6$.}\label{fig:sensor-noise}\end{figure}

{For OF, density and speed sensing at the downstream boundary reconstruct the collocated output. The density and speed could be obtained through cameras and radar, as well as loop-detectors. These sensors provide the necessary data for the observer, however, noise are inevitable in real-world applications. In this section, we model the effect of sensor noise on the observer performance and then test the the backstepping method under sensor noise. We model measured physical quantities as $\rho_m=\rho(L,t)+n_\rho$ and $v_m=v(L,t)+n_v$, where $n_\rho$ and $n_v$ are zero-mean Gaussian noise terms, then yielding
\begin{align}
y_m(t)=e^{L/(\tau v^*)}\left[v_m-v^*+\frac{S}{\rho^*}(\rho_m-\rho^*)\right].
\label{eq:physical-measurement}
\end{align}
The same $y_m$ replaces $y_1$ in observer injection and its downstream boundary. Measurement errors are not added directly to the plant's physical state. The plant is driven by $U_{b}=dU$, and the observer uses the same input. The actuator is assumed to apply the command exactly without magnitude and rate limits.

At $P=0.6$, independent zero-mean Gaussian errors are generated for the density and speed measurements at $1$ s intervals and linearly interpolated between sampling instants. Plant regulation uses $E_{\text{plant}}$ in~\eqref{eq:settling-norm} and observer error uses physical traffic states to define the observer error norm,
\begin{align}
E_{\text{obs}}(t)=\left\{\int_0^L\left[\left(\frac{\rho-\rho_{\rm est}}{\rho^*}\right)^2+
\left(\frac{v-v_{\rm est}}{v^*}\right)^2\right]dx\right\}^{1/2},
\label{eq:observer-error-norm}
\end{align}
where $\rho_{\rm est}$ and $v_{\rm est}$ are the estimated density and speed from the observer, respectively. Figure~\ref{fig:sensor-noise} shows the system norm and observer error norm under different noise levels. In the tested $0$--$400$ s runs, the initial perturbation declines and nonzero-noise cases retain fluctuations, with larger residuals at higher noise levels. No sustained growth is observed in this horizon, but none of the nonzero levels reaches and remains below the settling threshold by $400$ s. It shows that the observer is sensitive to sensor noise.The results show that the backstepping design may require additional filtering or robust design to improve performance.
}

\subsection{Numerical Evaluation with Real Traffic Data}\label{sec:calibration}
\begin{figure}[tbp]\centering
\includegraphics[width=.45\linewidth]{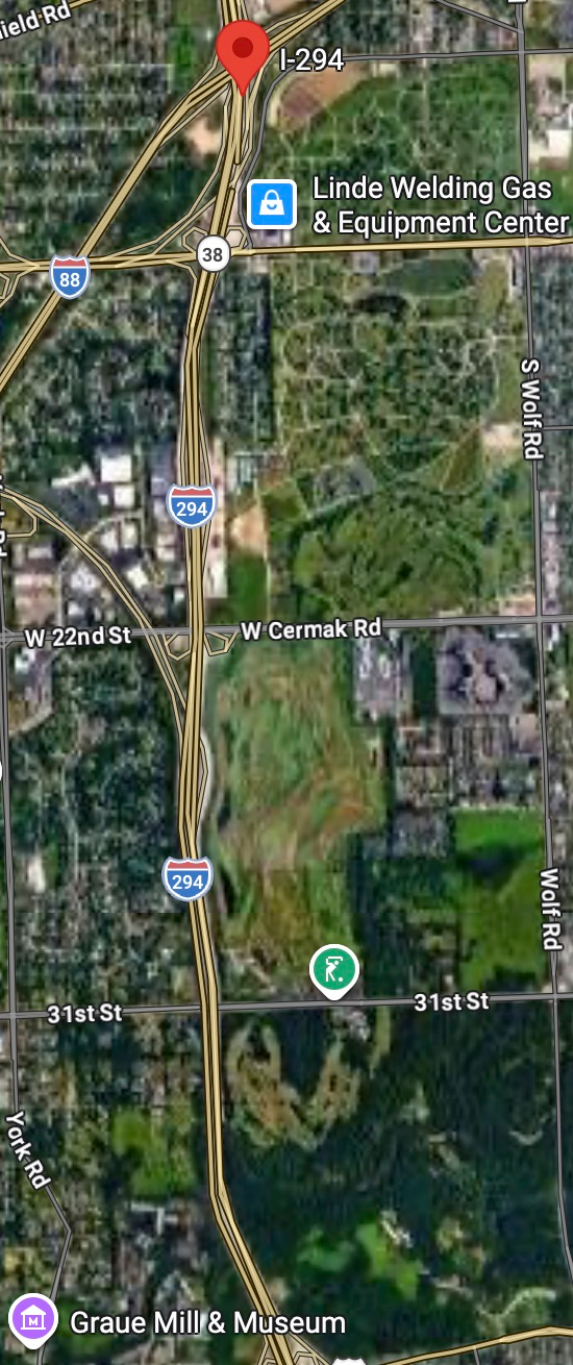}
\caption{I-294 collection location. Basemap: Google Maps~\cite{GoogleMaps2026}; data-site description~\cite{ammourah2025introduction}.}\label{I294seg}\end{figure}

We further test the controllers using a fundamental diagram calibrated from the Third Generation Simulation (TGSIM) trajectory data~\cite{ammourah2025introduction} to determine whether the proposed boundary control structure retains its stabilizing effect when the analytical mixed fundamental diagram is replaced by a diagram fitted to real traffic data. The TGSIM data were collected to study human interactions with vehicles equipped with driving automation in real traffic. The I-294 data used here were recorded between 15:00 and 17:00 near Hinsdale, Illinois, in October 2021, as shown in Fig.~\ref{I294seg}. The data were obtained using aerial videography: a helicopter followed a string of three SAE Level 1 ACC-driven vehicles while an 8K camera recorded vehicle trajectories at 30 fps from an altitude of about 300~m. The processed trajectories provide position, lane, speed, acceleration, length and ACC labels. The ACC-labeled subset contains $30$ of $1165$ run/vehicle identifiers ($2.58\%$). This is a trajectory-level statistic, not necessarily a unique physical fleet share or measured cooperative-CAV penetration. ACC-labeled vehicles supply a limited automated-following proxy, its limitations are discussed in Section~\ref{sec:discussion}.

{Following Edie's definitions~\cite{edie1963discussion}, vehicle trajectoreis are split across $15$ s by $20$ m time-space cells. For each retained cell, flow and density are computed from the total vehicle-distance and vehicle-time, respectively, normalized by the cell's area and by the fixed number of four southbound lanes. Related data-based estimation procedures for traffic density and speed are described in~\cite{zhao2023observer,fan2013data,YuPDETraffic2021}. In the normal-policy scenario, we use $v_f=55$ mph $=88.514$ km/h~\cite{IllinoisTollwaySpeed2015}. The mean per-trajectory vehicle length $8.167$ m, and with a prescribed $2$ m gap, this gives $\rho_j=1000/(8.167+2)=98.356$ veh/km/lane. These two quantities are fixed, and only the backward wave speed is fitted by bounded Huber-loss least squares with scale $100$ veh/h/lane and bounds $1$--$80$ km/h. After excluding cross-lane trajectory segments and applying the existing coverage-proxy screening, all valid cells are used for fitting. The resulting fit is
\begin{align}
Q(\rho)=\min\{88.514\rho,15.940(98.356-\rho)\},
\label{eq:calibrated-fd}
\end{align}
with $\rho_c=15.009$ veh/km/lane and capacity $1328.513$ veh/h/lane. Robust scatter regression has no upper-envelope constraint, so its peak need not pass through the largest observations. 
\begin{figure}[tbp]\centering
\subfloat[Flow--density]{\includegraphics[width=.48\linewidth]{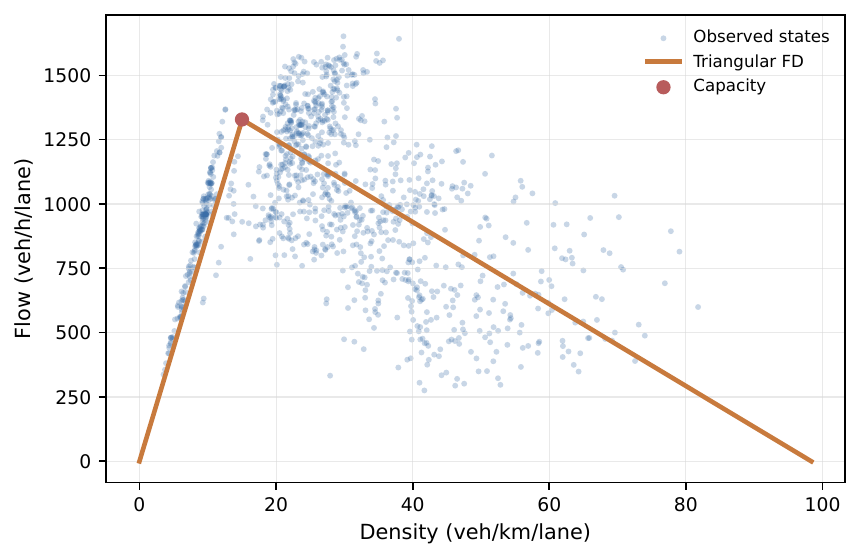}}
\subfloat[Speed--density]{\includegraphics[width=.48\linewidth]{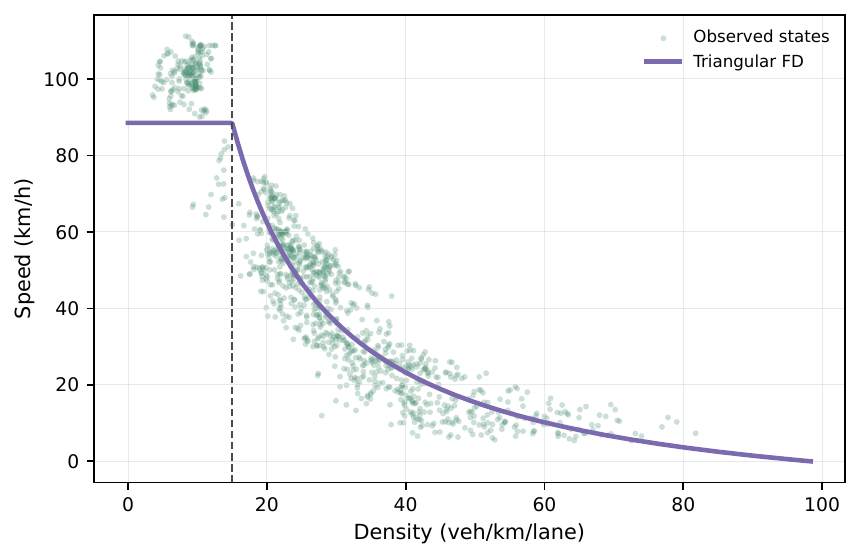}}
\caption{Calibrated fundamental diagram based on real traffic data.}\label{calibrated fd}\end{figure}

With $\rho^*=45$ veh/km/lane, the calibrated congested equilibrium has $v^*=18.899$ km/h. Then we apply the calibrated fundamental diagram to the same linearized SF/OF design, input map and initial conditions. Figures~\ref{real data control-sf} and~\ref{real data control-of} show the resulting traffic density and speed evolutions over the 400s simulation horizon. Both SF and OF converge to the calibrated equilibrium over this period. These results show that the data-based backstepping design retains its stabilizing effect when the analytical mixed fundamental diagram is replaced by a diagram calibrated from real traffic data, which supports the practical applicability of the proposed approach.
\begin{figure}[tbp]\centering
\subfloat[Density]{\includegraphics[width=.48\linewidth]{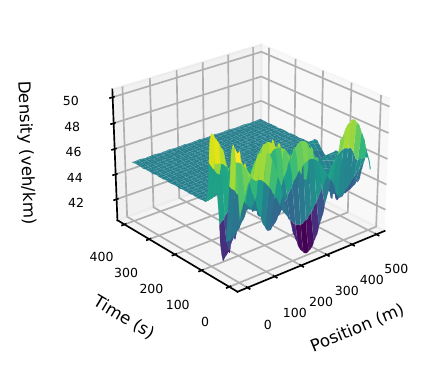}}
\subfloat[Speed]{\includegraphics[width=.48\linewidth]{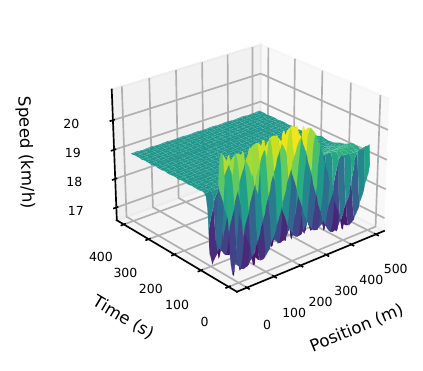}}
\caption{Physical SF states in the calibrated linearized model.}\label{real data control-sf}\end{figure}
\begin{figure}[!tbp]\centering
\subfloat[Density]{\includegraphics[width=.48\linewidth]{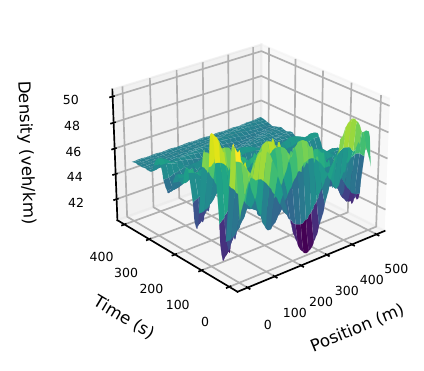}}
\subfloat[Speed]{\includegraphics[width=.48\linewidth]{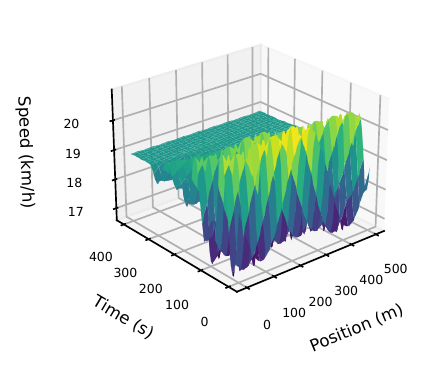}}
\caption{Physical OF plant states in the calibrated linearized model.}\label{real data control-of}\end{figure}
}

{

\section{Discussion and Limitation}\label{sec:discussion}
\textbf{Road geometry and multi-lane extensions.}
The numerical simulation is assumed a fixed $500$ m single-lane segment with one downstream regulation channel and no internal on-ramps or off-ramps. HV/CAV interactions enter through aggregate spacing and relaxation rather than separate vehicle-class dynamics. Internal ramp sources require additional balance laws and junction conditions. A multi-lane model must describe lane-specific densities, speeds and, where necessary, class composition, together with lane-change exchange terms that conserve the total vehicle and class counts. These couplings change wave propagation while the present $2\times2$ kernel equations and stability proof can not apply directly. Existing two-lane boundary-control results~\cite{YuOutputFeedback2021} provide preliminary insights, while extension and validation for this mixed diagram remain future work.

\textbf{Penetration uncertainty and adaptive estimation.}
The controller uses a known, uniform and constant penetration rate $P$. If its estimate $\widehat P$ differs from the actual value, the mixed headway and, when the class gaps differ, effective spacing change. This affects the reference equilibrium, characteristic speeds, control/observer kernels and output reconstruction. Reference mismatch may create a steady offset, while imperfect nominal cancellation may slow convergence or affect stability under larger errors. Online estimation and adaptive updates of the PDE controller and observer are a future direction based on adaptive control of hyperbolic PDEs~\cite{anfinsen2019adaptive}.

\textbf{ACC labels and CAV extrapolation.}
The I-294 sample contains $2.58\%$ ACC-labeled trajectories. SAE Level~1 ACC is considered as a proxy of CAV in this study due to the lack of CAV dataset, however, it does not fully reflect the driving behavior of true CAVs. ACC-labled vehicles and cooperative CAVs may differ in selected headways, acceleration responses and communication capabilities. The CAV assumption uses literature-informed scenario assumptions, with sensitivity quantified within the analytical model. More realistic CAV data will be needed to validate the proposed penetration-dependent model and associated control design.

\textbf{Non-uniform CAV distribution.}
We assume a uniform CAV distribution along the road segment in this study. In practice, CAVs and HVs may be clustered in certain regions, leading to non-uniform penetration. With equal gaps, CAV-rich regions have shorter headway and higher local capacity and equilibrium speed at a fixed congested density. Free speed and jam density are unchanged, while critical density and the congested slope differ. CAV-rich upstream supply may overload an HV-rich downstream region,  while upstream HV-rich supply may limit flow. At low demand, neither contrast necessarily creates congestion.

A frozen pure-CAV/pure-HV profile illustrates the need for coupled road-section models. For an interface $x_c$, let $P_1=1$, $P_2=0$ (or reverse them) deonte the penetration rate of CAVs of two road sections, and $V_j(\rho)=V_{\rm mix}(\rho;P_j)$. Each section obeys
\begin{align}
\begin{aligned}
\partial_t\rho_j+\partial_x(\rho_jv_j)&=0,\\
\partial_t w_j+v_j\partial_xw_j&=-w_j/\tau,\\
w_j&=v_j-V_j(\rho_j),\quad j=1,2.
\end{aligned}\label{eq:two-section-arz}
\end{align}
Without interface storage or sources, mass conservation requires
\begin{align}
(\rho_1v_1)(x_c^-,t)=(\rho_2v_2)(x_c^+,t).
\label{eq:interface-mass}
\end{align}
Flow continuity alone is insufficient to couple the two ARZ models. An additional interface condition specifying how $w_j$ is transmitted between the sections is required~\cite{HuWang2020}. Local equilibrium, propagation and transmission differences may impair a controller designed at uniform $P$. Actual composition regions move with vehicles and require class-conservation equations or a moving-interface description.
}

\section{Conclusion}\label{con}
This paper embeds a penetration-dependent fundamental diagram in an ARZ model and develops full-state and observer-based boundary controller. For the linearized congested system with exact constant parameters, ideal input and noise-free sensing, bounded coordinate and Volterra transformations yield exponential stability and finite-time convergence. Numerical simulations illustrate the stabilizing effect of the proposed controllers, comparing their performance with that of a downstream PI controller. The sensitivity of the closed-loop system, VSL magnitude and rate limits, and physical boundary sensor noise are also studied to demonstrate the robustness of the proposed controllers. We further test the controllers using a fundamental diagram calibrated from the TGSIM dataset, demonstrating the application potential for real-world traffic management. Future work includes adaptive penetration rate estimation, richer vehicle-class and multi-lane models, and analysis of the resulting coupled control systems.

\bibliography{references}

\begin{thebibliography}{10}

\bibitem{siri2021freeway}
S.~Siri, C.~Pasquale, S.~Sacone, and A.~Ferrara, ``{Freeway traffic control: A
  survey},'' {\em Automatica}, vol.~130, p.~109655, 2021.

\bibitem{treiber2013traffic}
M.~Treiber and A.~Kesting, {\em Traffic flow dynamics: data, models and
  simulation}.
\newblock Springer Science \& Business Media, 2012.

\bibitem{papageorgiou1991alinea}
M.~Papageorgiou, H.~Hadj-Salem, J.-M. Blosseville, {\em et~al.}, ``{ALINEA: A
  local feedback control law for on-ramp metering},'' {\em Transportation
  Research Record}, vol.~1320, no.~1, pp.~58--67, 1991.

\bibitem{karafyllis2019feedback}
I.~Karafyllis and M.~Papageorgiou, ``{Feedback control of scalar conservation
  laws with application to density control in freeways by means of variable
  speed limits},'' {\em Automatica}, vol.~105, pp.~228--236, 2019.

\bibitem{zhang2017stochastic}
L.~Zhang and C.~Prieur, ``{Stochastic stability of Markov jump hyperbolic
  systems with application to traffic flow control},'' {\em Automatica},
  vol.~86, pp.~29--37, 2017.

\bibitem{gomes2006optimal}
G.~Gomes and R.~Horowitz, ``{Optimal freeway ramp metering using the asymmetric
  cell transmission model},'' {\em Transportation Research Part C: Emerging
  Technologies}, vol.~14, no.~4, pp.~244--262, 2006.

\bibitem{goatin2016speed}
P.~Goatin, S.~G{\"o}ttlich, and O.~Kolb, ``Speed limit and ramp meter control
  for traffic flow networks,'' {\em Engineering Optimization}, vol.~48, no.~7,
  pp.~1121--1144, 2016.

\bibitem{delle2017traffic}
M.~L. Delle~Monache, B.~Piccoli, and F.~Rossi, ``{Traffic regulation via
  controlled speed limit},'' {\em SIAM Journal on Control and Optimization},
  vol.~55, no.~5, pp.~2936--2958, 2017.

\bibitem{belletti2017expert}
F.~Belletti, D.~Haziza, G.~Gomes, and A.~M. Bayen, ``{Expert level control of
  ramp metering based on multi-task deep reinforcement learning},'' {\em IEEE
  Transactions on Intelligent Transportation Systems}, vol.~19, no.~4,
  pp.~1198--1207, 2017.

\bibitem{wang2025adaptive}
Z.~Wang, Y.~Zhang, C.~Zhao, and H.~Yu, ``Adaptive event-triggered formation
  control of autonomous vehicles,'' {\em arXiv preprint arXiv:2506.06746},
  2025.

\bibitem{wang2026event}
Z.~Wang, X.~Ge, and Q.-L. Han, ``Event-triggered v2x communication and control
  co-design for intelligent electrified transportation systems,'' {\em Journal
  of Intelligent and Connected Vehicles}, vol.~9, no.~3, 2026.

\bibitem{wang2026safe}
Z.~Wang, S.~Piao, X.~Wang, F.~Tsung, K.~H. Johansson, and A.~I. Rikos, ``Safe:
  Switched adaptive formation via event-triggered control for multi-vehicle
  systems,'' {\em IEEE Systems, Man, and Cybernetics Letters}, vol.~1,
  pp.~100--107, 2026.

\bibitem{yu2018varying}
H.~Yu and M.~Krstic, ``Varying speed limit control of {Aw-Rascle-Zhang} traffic
  model,'' in {\em 2018 21st international conference on intelligent
  transportation systems (ITSC)}, pp.~1846--1851, IEEE, 2018.

\bibitem{papageorgiou2002freeway}
M.~Papageorgiou and A.~Kotsialos, ``Freeway ramp metering: An overview,'' {\em
  IEEE transactions on Intelligent Transportation Systems}, vol.~3, no.~4,
  pp.~271--281, 2002.

\bibitem{TalebpourInfluenceConnected2016}
A.~Talebpour and H.~S. Mahmassani, ``Influence of connected and autonomous
  vehicles on traffic flow stability and throughput,'' {\em Transportation
  Research Part C: Emerging Technologies}, vol.~71, pp.~143--163, 2016.

\bibitem{SternDissipationStopandgo2018}
R.~E. Stern, S.~Cui, M.~L. Delle~Monache, R.~Bhadani, M.~Bunting, M.~Churchill,
  N.~Hamilton, R.~Haulcy, H.~Pohlmann, F.~Wu, B.~Piccoli, B.~Seibold,
  J.~Sprinkle, and D.~B. Work, ``Dissipation of stop-and-go waves via control
  of autonomous vehicles: Field experiments,'' {\em Transportation Research
  Part C: Emerging Technologies}, vol.~89, pp.~205--221, 2018.

\bibitem{lighthill1955kinematic}
M.~J. Lighthill and G.~B. Whitham, ``On kinematic waves ii. a theory of traffic
  flow on long crowded roads,'' {\em Proceedings of The Royal Society of
  London. Series A. Mathematical and Physical Sciences}, vol.~229, no.~1178,
  pp.~317--345, 1955.

\bibitem{richards1956shock}
P.~I. Richards, ``Shock waves on the highway,'' {\em Operations Research},
  vol.~4, no.~1, pp.~42--51, 1956.

\bibitem{payne1971model}
H.~J. Payne, ``Model of freeway traffic and control,'' {\em Mathematical Model
  of Public System}, pp.~51--61, 1971.

\bibitem{whitham2011linear}
G.~B. Whitham, {\em Linear and nonlinear waves}.
\newblock John Wiley \& Sons, 2011.

\bibitem{aw2000resurrection}
A.~Aw and M.~Rascle, ``Resurrection of" second order" models of traffic flow,''
  {\em SIAM Journal on Applied Mathematics}, vol.~60, no.~3, pp.~916--938,
  2000.

\bibitem{zhang2002non}
H.~M. Zhang, ``A non-equilibrium traffic model devoid of gas-like behavior,''
  {\em Transportation Research Part B: Methodological}, vol.~36, no.~3,
  pp.~275--290, 2002.

\bibitem{yu2019traffic}
H.~Yu and M.~Krstic, ``Traffic congestion control for {Aw--Rascle--Zhang}
  model,'' {\em Automatica}, vol.~100, pp.~38--51, 2019.

\bibitem{ghiasi2017mixed}
A.~Ghiasi, O.~Hussain, Z.~S. Qian, and X.~Li, ``A mixed traffic capacity
  analysis and lane management model for connected automated vehicles: A
  {Markov} chain method,'' {\em Transportation Research Part B:
  Methodological}, vol.~106, pp.~266--292, 2017.

\bibitem{shi2021constructing}
X.~Shi and X.~Li, ``Constructing a fundamental diagram for traffic flow with
  automated vehicles: Methodology and demonstration,'' {\em Transportation
  Research Part B: Methodological}, vol.~150, pp.~279--292, 2021.

\bibitem{yao2019stability}
Z.~Yao, R.~Hu, Y.~Wang, Y.~Jiang, B.~Ran, and Y.~Chen, ``Stability analysis and
  the fundamental diagram for mixed connected automated and human-driven
  vehicles,'' {\em Physica A: Statistical Mechanics and Its Applications},
  vol.~533, p.~121931, 2019.

\bibitem{zhou2020modeling}
J.~Zhou and F.~Zhu, ``Modeling the fundamental diagram of mixed human-driven
  and connected automated vehicles,'' {\em Transportation Research Part C:
  Emerging Technologies}, vol.~115, p.~102614, 2020.

\bibitem{ye2018modeling}
L.~Ye and T.~Yamamoto, ``Modeling connected and autonomous vehicles in
  heterogeneous traffic flow,'' {\em Physica A: Statistical Mechanics and Its
  Applications}, vol.~490, pp.~269--277, 2018.

\bibitem{chang2020analysis}
X.~Chang, H.~Li, J.~Rong, X.~Zhao, {\em et~al.}, ``Analysis on traffic
  stability and capacity for mixed traffic flow with platoons of intelligent
  connected vehicles,'' {\em Physica A: Statistical Mechanics and Its
  Applications}, vol.~557, p.~124829, 2020.

\bibitem{BurkhardtStopandgoSuppression2021}
M.~Burkhardt, H.~Yu, and M.~Krstic, ``Stop-and-go suppression in two-class
  congested traffic,'' {\em Automatica}, vol.~125, p.~109381, 2021.

\bibitem{ZhangMeansquareExponential2024}
Y.~Zhang, H.~Yu, J.~Auriol, and M.~Pereira, ``Mean-square exponential
  stabilization of mixed-autonomy traffic {PDE} system,'' {\em Automatica},
  vol.~170, p.~111859, 2024.

\bibitem{ZhangEventTriggeredRegulation2026}
Y.~Zhang and H.~Yu, ``Event-triggered regulation of mixed-autonomy traffic
  under varying traffic conditions,'' {\em IEEE Transactions on Intelligent
  Transportation Systems}, vol.~27, no.~1, pp.~816--830, 2026.

\bibitem{zhang2024event}
Y.~Zhang and H.~Yu, ``Event-triggered boundary control of mixed-autonomy
  traffic,'' in {\em 2024 IEEE 63rd Conference on Decision and Control (CDC)},
  pp.~6428--6433, IEEE, 2024.

\bibitem{krstic2008boundary}
M.~Krstic and A.~Smyshlyaev, {\em Boundary control of {PDEs}: A course on
  backstepping designs}.
\newblock SIAM, 2008.

\bibitem{VazquezBacksteppingBoundary2011}
R.~Vazquez, M.~Krstic, and J.-M. Coron, ``Backstepping boundary stabilization
  and state estimation of a 2 × 2 linear hyperbolic system,'' in {\em 2011
  50th IEEE Conference on Decision and Control and European Control
  Conference}, pp.~4937--4942, 2011.

\bibitem{anfinsen2019adaptive}
H.~Anfinsen and O.~M. Aamo, {\em Adaptive control of hyperbolic PDEs}.
\newblock Springer, 2019.

\bibitem{YuOutputFeedback2021}
H.~Yu and M.~Krstic, ``Output feedback control of two-lane traffic
  congestion,'' {\em Automatica}, vol.~125, p.~109379, 2021.

\bibitem{bekiaris2020pde}
N.~Bekiaris-Liberis and A.~I. Delis, ``{PDE}-based feedback control of freeway
  traffic flow via time-gap manipulation of {ACC}-equipped vehicles,'' {\em
  IEEE Transactions on Control Systems Technology}, vol.~29, no.~1,
  pp.~461--469, 2020.

\bibitem{qi2022delay}
J.~Qi, S.~Mo, and M.~Krstic, ``Delay-compensated distributed {PDE} control of
  traffic with connected/automated vehicles,'' {\em IEEE Transactions on
  Automatic Control}, 2022.

\bibitem{yao2023analysis}
Z.~Yao, Y.~Wu, Y.~Wang, B.~Zhao, and Y.~Jiang, ``Analysis of the impact of
  maximum platoon size of {CAVs} on mixed traffic flow: An analytical and
  simulation method,'' {\em Transportation Research Part C: Emerging
  Technologies}, vol.~147, p.~103989, 2023.

\bibitem{kesting2010enhanced}
A.~Kesting, M.~Treiber, and D.~Helbing, ``Enhanced intelligent driver model to
  access the impact of driving strategies on traffic capacity,'' {\em
  Philosophical Transactions of The Royal Society A: Mathematical, Physical and
  Engineering Sciences}, vol.~368, no.~1928, pp.~4585--4605, 2010.

\bibitem{treiber2000congested}
M.~Treiber, A.~Hennecke, and D.~Helbing, ``Congested traffic states in
  empirical observations and microscopic simulations,'' {\em Physical review
  E}, vol.~62, no.~2, pp.~1805--1824, 2000.

\bibitem{milanes2013cooperative}
V.~Milan{\'e}s, S.~E. Shladover, J.~Spring, C.~Nowakowski, H.~Kawazoe, and
  M.~Nakamura, ``Cooperative adaptive cruise control in real traffic
  situations,'' {\em IEEE Transactions on Intelligent Transportation Systems},
  vol.~15, no.~1, pp.~296--305, 2014.

\bibitem{ZhangPI2019}
L.~Zhang, C.~Prieur, and J.~Qiao, ``{PI} boundary control of linear hyperbolic
  balance laws with stabilization of {ARZ} traffic flow models,'' {\em Systems
  \& Control Letters}, vol.~123, pp.~85--91, 2019.

\bibitem{Shreim2022}
S.~Shreim, F.~Ferrante, and C.~Prieur, ``Design of saturated boundary control
  for hyperbolic systems with in-domain disturbances,'' {\em Automatica},
  vol.~142, p.~110346, 2022.

\bibitem{GomesDaSilva2003}
J.~G. Da~Silva, S.~Tarbouriech, and G.~Garcia, ``Local stabilization of linear
  systems under amplitude and rate saturating actuators,'' {\em IEEE
  Transactions on Automatic Control}, vol.~48, no.~5, pp.~842--847, 2003.

\bibitem{ammourah2025introduction}
R.~Ammourah, P.~Beigi, B.~Fan, S.~H. Hamdar, J.~Hourdos, C.-C. Hsiao, R.~James,
  M.~Khajeh-Hosseini, H.~S. Mahmassani, D.~Monzer, {\em et~al.}, ``Introduction
  to the third generation simulation dataset: Data collection and trajectory
  extraction,'' {\em Transportation Research Record}, vol.~2679, no.~1,
  pp.~1768--1784, 2025.

\bibitem{GoogleMaps2026}
{Google}, ``{Google Maps}.'' \url{https://www.google.com/maps}, 2026.
\newblock Accessed: 2026-10-11.

\bibitem{edie1963discussion}
L.~C. Edie, ``Discussion of traffic stream measurements and definitions,'' in
  {\em Proceedings of the Second International Symposium on the Theory of
  Traffic Flow}, pp.~139--154, Organisation for Economic Co-operation and
  Development, 1965.

\bibitem{zhao2023observer}
C.~Zhao and H.~Yu, ``Observer-informed deep learning for traffic state
  estimation with boundary sensing,'' {\em IEEE Transactions on Intelligent
  Transportation Systems}, vol.~25, no.~2, pp.~1602--1611, 2024.

\bibitem{fan2013data}
S.~Fan and B.~Seibold, ``Data-fitted first-order traffic models and their
  second-order generalizations: Comparison by trajectory and sensor data,''
  {\em Transportation Research Record}, vol.~2391, no.~1, pp.~32--43, 2013.

\bibitem{YuPDETraffic2021}
H.~Yu, Q.~Gan, A.~Bayen, and M.~Krstic, ``{PDE} traffic observer validated on
  freeway data,'' {\em IEEE Transactions on Control Systems Technology},
  vol.~29, no.~3, pp.~1048--1060, 2021.

\bibitem{IllinoisTollwaySpeed2015}
{Illinois Tollway}, ``Tollway speed limit recommendations,'' tech. rep.,
  Illinois Tollway, March 2015.

\bibitem{HuWang2020}
Y.~Hu and G.~Wang, ``The resonance behavior for the coupling of two
  {Aw--Rascle} traffic models,'' {\em Advances in Difference Equations},
  vol.~2020, p.~60, 2020.

\end{thebibliography}
\bibliographystyle{ieeetr}

\end{document}